\documentclass{aa} 
\usepackage{booktabs}
\usepackage{graphicx}
\usepackage{txfonts}
\usepackage{multirow}
\usepackage{xcolor}
\begin{document}

 \title{Two-horn {quiescent} 
 prominence observed in H$\alpha$ and \ion{Mg}{ii}~h\&k lines with THEMIS and IRIS}

\titlerunning{Two-horn prominence with THEMIS and IRIS}
\authorrunning{Barczynski et al.}

 \author{Krzysztof Barczynski \inst{1,2},
Brigitte Schmieder\inst{3,4,5},
Bernard Gelly\inst{6},
Aaron W. Peat\inst{4,7},
Nicolas Labrosse\inst{4}
 }

\institute{ETH-Zurich, H\"onggerberg campus, HIT building, Z\"urich, Switzerland
\\
 \email{krzysztof.barczynski@pmodwrc.ch}
 \and
 PMOD/WRC, Dorfstrasse 33, CH-7260 Davos Dorf, Switzerland
\and
LESIA, Observatoire de Paris, Universit\'e PSL , CNRS, Sorbonne Universit\'e, Universit\'e Paris-Diderot, 5 place Jules Janssen, 92190 Meudon, France 
\and
SUPA School of Physics and Astronomy, The University of Glasgow, Glasgow, G12 8QQ, UK
\and
KU-Leuven, 3001 Leuven, Belgium
\and
THEMIS, INSU,CNRS,UPS 3718 IAC, Via Lactea, 38200 La Laguna, Tenerife, Spain
\and
University of Wroc\l{}aw, Centre of Scientific Excellence -- Solar and Stellar Activity, Kopernika 11, 51-622 Wroc\l{}aw, Poland
}

 \date{Received xxx, xxx; accepted xxx, xxx}

 \abstract
 {Prominences are large magnetic structures in the corona filled by cool plasma with fast evolving fine structure.}
 {We aim to better understand the plasma conditions in the fine structure of {a quiescent prominence} including two transient horns observed{ at the bottom of the cavity} using the high resolution Interface Region Imaging Spectrograph (IRIS) and the MulTi-Raies (MTR) spectrograph {of the T\'{e}lescope Heliographique pour l'Etude du Magn\'{e}tisme et des Instabilit\'{e}s Solaires (THEMIS)} in the Canary Islands.}
 {We analysed the spectra obtained in H$\alpha$ {by THEMIS} and \ion{Mg}{ii} {by IRIS and compare them with} a grid of 23\,940 1D radiative transfer models which include {a} prominence-to-corona transition region (PCTR). The full observed profiles of \ion{Mg}{ii} in each pixel are fitted completely by synthesised profiles with {xRMS (Cross RMS; an improved version of the rolling root mean square (rRMS) method)}. When the RMS is below {a certain threshold value, we recover} the plasma conditions from the parameters of the model best fitting the observed line profile. 
 { This criterion is met in two regions (the horns and edge of the prominence) where the line profiles can generally be described as single peaked.
 }}
 {{
 The 1D models suggest that two different kinds of model atmospheres correspond to these two regions. The region at the edge is found to be fitted mainly with isothermal and isobaric models, while  the other area ({the} horns) is seen to be {fitted with} models with a PCTR {that have} optical thicknesses of less than 5. In the prominence edge, the theoretical relationship between the integrated intensities in H$\alpha$ and \ion{Mg}{ii} is verified and corresponds to low emission measure values. In these regions the electron density is around 10$^{10}$~cm$^{-3}$ , while it is one order of magnitude less in the horn regions around 10$^9$~cm$^{-3}$. }}
 {In the horns, we find some profiles are best fitted with models with high mean temperatures. This suggests 
that {the hot PCTR found in the} horns could be interpreted as prominence plasma in condensation phase at the bottom of the coronal cavity.}
 \keywords{Sun: prominence --- Sun: magnetic fields}
 
 \maketitle
%
\section{Introduction} 
\label{sec:intro}
Prominences are intriguing and enigmatic structures with a fine balance of forces ({e.g. gravity and magnetic tension), which exhibit} evolving magnetic topology.
Solar prominences are cool dense structures suspended in the solar corona,
at heights typically ranging from a few thousand kilometres up to 
hundreds of thousands of kilometres. 
Typical temperatures of the central regions of prominences are between 6000 and 8500~K. Outside of
this region, the temperature and pressure rapidly increase up to coronal values of around 1~MK and{ 0.05~dyn~cm$^{-2}$ }\citep{aschwanden_physics_2004}. The region where this rapid temperature and pressure change occurs is called the prominence-to-corona transition region (PCTR). The plasma density in the central coolest
parts is about two orders of magnitude larger than that in the corona, and thus
the presence of magnetic field is necessary for the support and stability of a prominence. Nevertheless, it is surprising that this relatively cool material reaches coronal heights. The driving forces behind their stability, evolution, and fine structure remain open questions in the field \citep{Schmieder2004b}. 

Once formed, prominences come in a wide variety of {morphologies}. They display large variet{ies of} dynamics, which are understood to be driven by their formation, and internal and global evolution \citep{Labrosse2010,Mackay2010}. { Their appearance is strongly influenced by projection effects \citep{Schmieder2017,Gunar2018}.}

\begin{figure*}[!htb]
\centering
\includegraphics[width=\textwidth]{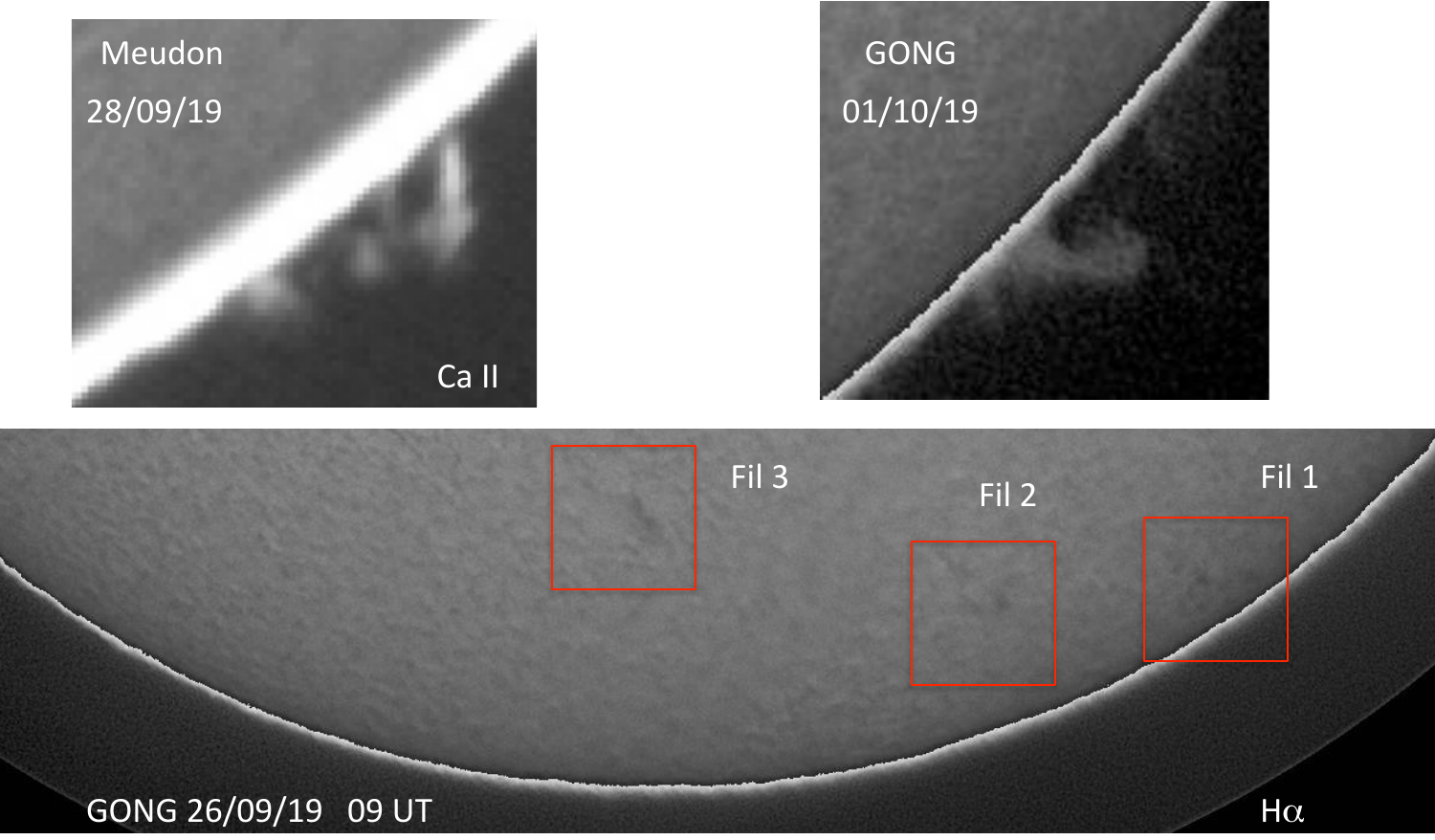}
\caption{Filaments and prominences in survey images from Meudon spectroheliograms in Ca II and from GONG in H$\alpha$. The prominence on September 28 corresponds to filament F1.}
\label{GONG}
\end{figure*}

\begin{figure*}[!htb]
\centering
\includegraphics[width=\textwidth]{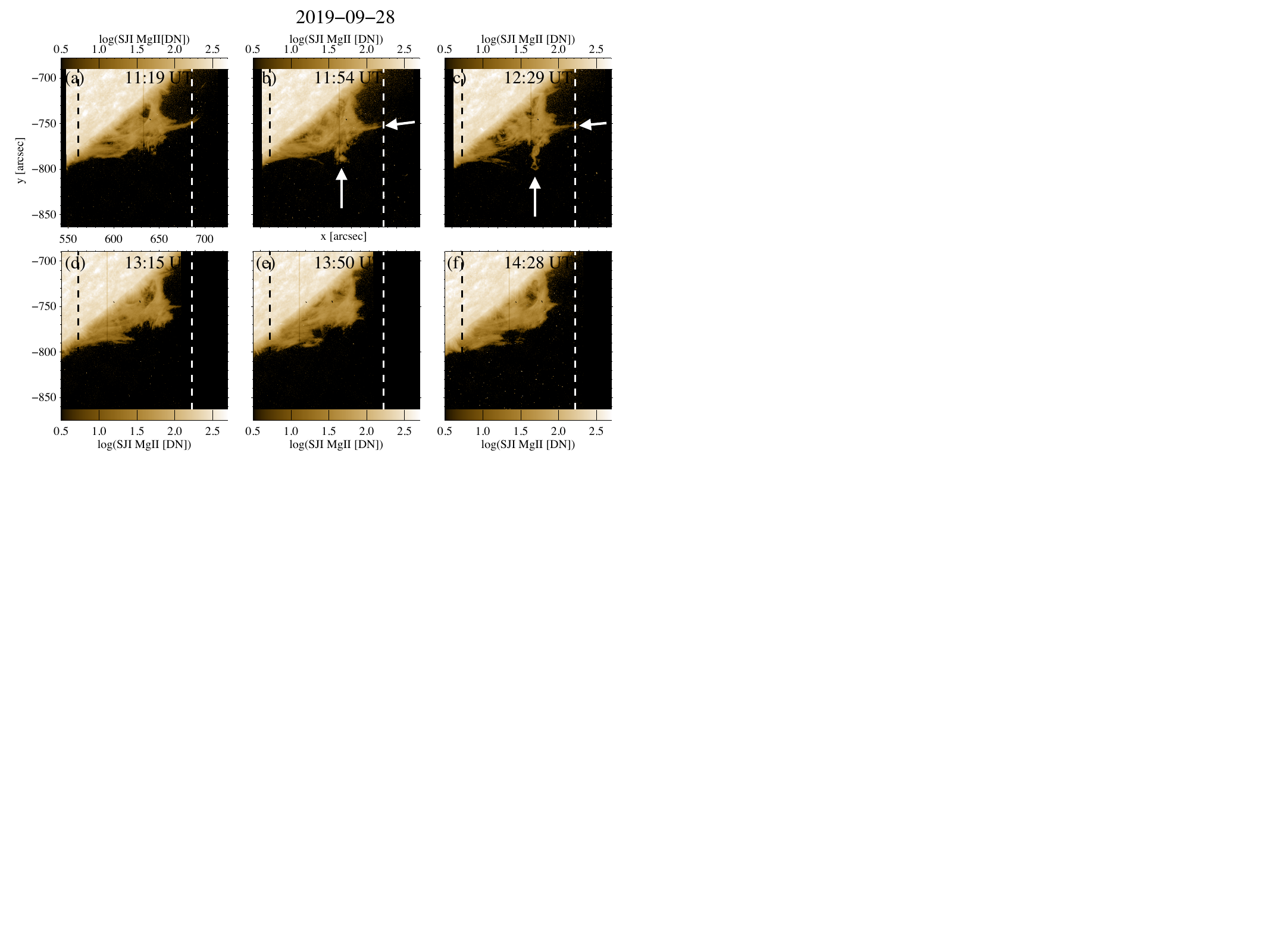}
\caption{Overview of the prominence in IRIS SJI images in \ion{Mg}{ii} during the three hours of observations, including snapshots of the movie (SJI2796.mp4). The vertical dashed lines indicate the field of view of the rasters. The two horns of the prominence are well visible in the top panels; white arrows point to the location of the horns.} \label{SJI}
\end{figure*}

\begin{figure}[!htb]
\centering
\includegraphics[width=\linewidth]{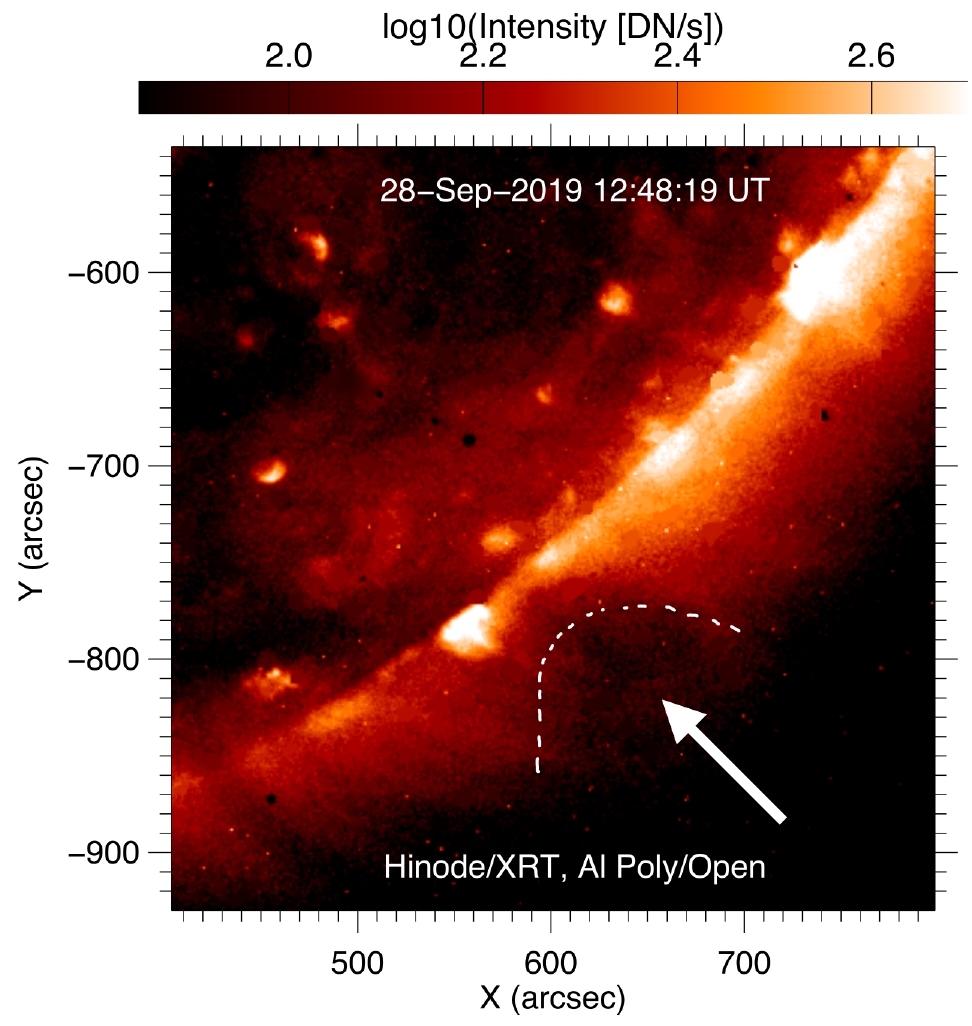}
\caption{Hinode/XRT image clearing showing the cavity around 620\arcsec, -800\arcsec { (half ellipse indicated by a white arrow}).} \label{XRT}
\end{figure}

{ Prominences may appear at the limb surrounded by a cavity when their corresponding filaments have their axis lying along the east-west direction{ (\citet{Gibson2018})}.{ \citet{Gibson2015} observed well contrasted cavities in AIA 171~\AA{} and 211~\AA. } A cavity is a tunnel-like plasma depletion with a coronal temperature and a central hotter core. The cavity reflects the fact that prominences have a flux rope topology with dips in the magnetic field lines \citep{Aulanier2002}.} 

Horns connect the cool and dense prominence core to the hot and tenuous corona, representing a field-aligned prominence-corona transition region where large amplitude oscillations may appear \citep{Luna2012,Wang2016}. Horns are located at the base of coronal cavities and have different plasma characteristics to that of the prominence body \citep{Schmit2013a,Schmit2013b}. These authors proposed that horns are formed by cooling and condensating plasma. This is supported by magnetohydrodynamic (MHD) simulations \citep{Luna2012,Schmit2013b,Xia2014}. On the contrary,
\citet{Wang2016} described the horns as the sites of heating and diluted plasma. From the MHD simulations of 
\citet{Fan2019}, horns are described as the sites of high current concentrations among quasi-separatrices and hyperbolic flux tube. Prominence `horns' are threaded by twisted field lines containing shallow dips, where the prominence condensations have evaporated to coronal temperatures \citep{Fan2019}. In horns, it is predicted that magnetic field lines are weakly twisted favouring high-speed flows and very transient condensation
\citep{GuoJH2022}.

{The modelling of solar prominences in order to obtain their diagnostics is a topic of current scientific research. As of now, the use of 1D NLTE modelling is the most computationally efficient due to the exponentially increasing parameter space of current 2D multithread models \citep{Gunar2022,Peat2023}.}
\citet{Jejcic2022} based their fitting on five parameters with isobaric and isothermal models -- the integrated intensity of H$\alpha$, \ion{Mg}{ii}~h, \ion{Mg}{ii}~k, their FWHM, and the k/h ratio. These observations 
were also presented in \citet{Ruan2018,Ruan2019}, but a smaller number of combined profiles were analysed. Using a similar method of fitting observations with 1D isobaric model results, \citet{Vial2019} and \citet{Zhang2019} discussed the physical parameters pixel by pixel at the base of an eruptive prominence from IRIS spectra.

{ A step forward compared to this multi-parameter fitting method {is the} fitting of the full observed \ion{Mg}{ii}{~h\&k profiles. These profiles are fitted using 1D radiative transfer models. These models can use one of two types of atmospheric model; an isothermal and isobaric atmosphere, or an atmosphere with a PCTR. This method of fitting has the benefit of fitting
the intensity in each wavelength point of {the observed and synthesised profiles}. Additionally, this consequently recovers the integrated intensities {and} FWHM {of the profiles}. This is a useful step to determine the physical parameters of fine structure in prominences {which exhibit} mainly non-reversed profiles. 
 \citet{Heinzel2015} showed that single-peak profiles could correspond to a fine structure with a relatively high gas pressure in models with PCTR. {This is} contrary to isobaric models, which fit single-peaked profiles with only low gas pressures \citep{Heinzel2014}}.
 
 {From} this perspective, \citet{Peat2021} developed the rRMS method to match the output of the 1D NLTE radiative transfer code, PROM \citep{Gouutebroze1993, heinzel1994, Levens2019}, directly to that of observation. In this aforementioned paper, full \ion{Mg}{ii}~h\&k profiles were fitted from a grid consisting of 1007 pre-generated models.
For every raster, each of the synthesised line profiles -- produced from the 1007 models -- is `rolled' through some wavelength window with the RMS at each pixel position being measured. The model that produces the lowest RMS value is selected as the best fit for that pixel. Additionally, a Dopplershift can be recovered from this method as a consequence of the rolling. \citet{Peat2021} demonstrated that the Dopplershift map obtained from rRMS was consistent with the Dopplershifts derived {from} the quantile method \citep{Kerr2015}. This method has since been improved with a {vectorised} cross-correlation and now produces conservative errors associated with each of the fits \citep{peatprep}.

In this paper we present a highly dynamic, twin-horned{ quiescent} prominence, observed with high resolution spectrographs in H$\alpha$ and \ion{Mg}{ii} (Table \ref{tab:lines}). 
Section~\ref{sec:obs_and_prep} describes the observations and calibration; Section~\ref{sec:morphology} {covers} the morphology of the prominence; Section~\ref{sec:char_ha_mg2} {details} the characteristics of the observed profiles (H$\alpha$ and \ion{Mg}{ii}); Section~\ref{sec:grid} presents the theoretical grid of models; Section~\ref{sec:phys_par} {provides} the theoretical parameters that are deduced from the comparison {of} theory {and} observations; {and, finally,} Section~\ref{sec:dis_concl} summarises the results.\\

\section{Observation and data preparation}\label{sec:obs_and_prep}
\subsection{Observation}\label{sec:observation}

In this work, we used simultaneous observations obtained by the Interface Region Imaging Spectrograph \cite[IRIS;][]{DePontieu2014}, the T\'{e}lescope Heliographique pour l'Etude du Magn\'{e}tisme et des Instabilit\'{e}s Solaires \citep[THEMIS;][]{mein_themis_1985}, the X-Ray Telescope \citep[XRT;][]{golub_x-ray_2007} onboard Hinode \citep{kosugi_hinode_2007}, and the Atmospheric Imaging Assembly \citep[AIA;][]{Lemen2012} onboard the Solar Dynamics Observatory \citep[SDO;][]{Pesnell2012}.
We focused on prominences at the west solar limb and analysed their properties based on imaging and spectroscopic observations.
 
{On 26 September 2019, images from} the Global Oscillation Network Group \citep[GONG;][]{gong}, of the National Solar Observatory (NSO) show the presence of a long polar filament channel in the southern hemisphere along the S40 parallel. This channel displayed a few bushes of dark filamentary material, which can be seen in Figure~\ref{GONG}. The Meudon spectroheliograph of the Observatoire de Paris observed {this} prominence in \ion{Ca}{ii}~H\&K on 28 September. Later, a second prominence was seen in H$\alpha$ by GONG on 1 October. These prominences were the targets of THEMIS and IRIS on 28 September 2019 and 1 October 2019, respectively. In Table \ref{tab:lines}, we summarise the characteristics of the spectral lines and instrument filters used for {these} IRIS, THEMIS, and AIA {observations.}

\begin{table*}[]
\centering
\caption{Observed line or band characteristics.}
\label{tab:lines}
\begin{tabular}{@{}lllll@{}}
\toprule
Instrument & Wavelength & Line or band & log $T$ {[}K{]} & Atmospheric regime \\ \midrule
\multirow{3}{*}{IRIS} & 2796 & \ion{Mg}{ii}~k & 3.6 - 3.9 & Chromosphere \\
 & 2803 & \ion{Mg}{ii}~h & 3.6 - 3.9 & Chromosphere \\
 & & \ion{Mg}{ii}~k peak & 3.9 & Upper chromosphere \\ \midrule
{THEMIS} & & H$\alpha$ & 3.6 - 3.9 & Chromosphere \\
 \midrule
\multirow{3}{*}{SDO/AIA} & 304 & \ion{He}{ii} & 4.7 & Chromosphere, TR \\
 & 171 & \ion{Fe}{ix} & 5.5 & Upper transition region \\
 & 193 & \ion{Fe}{xii}, \ion{Fe}{xxiv} & 6 & Corona \\ \bottomrule 
\end{tabular}
\end{table*}

\subsection{IRIS}
IRIS can provide observations in two far-ultraviolet (FUV) channels (FUV1: 1332-1358\AA, and FUV2: 1389-1407\AA) and a single near-ultraviolet (NUV) channel (NUV: 2783-2834\AA). 
These channels include strong chromospheric (\ion{Mg}{ii}, \ion{C}{ii}) and transition region (\ion{Si}{iv}) lines.
The observation program of 28 September 2019 focused {only} on the \ion{Mg}{ii} lines. We used the raster data to provide {statistics of} the \ion{Mg}{ii} (intensity, Doppler velocity, and FWHM).
The IRIS Slit Jaw Imager (SJI) provides the context of the position and surroundings of the slit with the 2796~\AA{} filter, which has a bandpass of 4~\AA.

We focused on the simultaneous \ion{Mg}{ii} rasters and SJI 2796~\AA\ observations of the prominence obtained on 28 September 2019 between 11:09 UTC and 14:56 UTC.
There is only SJI 2796~\AA\ from slit-jaw data for this observation.

The observation comprised of thirteen very large, coarse, 64-step rasters taken over a duration of approximately four hours.
The spatial step of the rasters is 2".
It took 16 seconds to perform one single slit observation and 17 minutes to perform a full raster scan including the time of reading the device.
{Meanwhile, the cadence of} the SJI is 16 seconds.
Thus, we have one SJI image for each slit position.
The SJIs allow us to study the fast transverse dynamics of the prominence material, which we cannot do with the spectra.
The details of the IRIS observations are summarised in Table~\ref{tab:instr_overview}. The data were downloaded from the Lockheed Martin Solar and Astrophysics Laboratory (LMSAL) IRIS database\footnote{\url{https://iris.lmsal.com/search/}.}. We used IRIS level-2 data corrected for the dark current, flat field, and geometric distortion \citep{DePontieu2014}.

The IRIS data are given in Data Number units (DN). The conversion to a flux in physical units was provided using IDL SolarSoft \citep[SSW;][]{freeland1998} procedure iris\_getwindata\footnote{\url{https://www.heliodocs.com/xdoc/xdoc_print.php?file=$SSW/iris/idl/nrl/iris_getwindata.pro}} with keywords /calib and /perang. The obtained intensity is presented in erg s$^{-1}$ sr$^{-1}$ cm$^{-2}$ \AA$^{-1}$.

\subsection{THEMIS}
To study the H$\alpha$ prominence, we used the observations provided by THEMIS on the island of Tenerife in the Canary Islands. The observations have the following characteristics: a lateral
spatial resolution of 0.5\arcsec{} (1 pixel size= 0.234/0.250\arcsec), a spectral resolution of 11/12 m\AA\ per pixel with a bandpass of 6.3~\AA\ centred around 6563~\AA{}. The field of view 
(FOV) is 100\arcsec$\times$128\arcsec. The exposure time is 100~ms, taking around one minute to complete a full raster; {however, here,} with the accumulation of data during 1 sec, the cadence is around 2.5 minutes.

Two prominences were well observed with IRIS during this campaign; the first of which was observed
on the 28 September 2019 located on the south western solar limb at S44 between 
 11:15-15:19 UTC (sequences 64-121); and the second
 on the 1 October 2019 also located on the south western solar limb at S20 between 08:09-09:54 and 10:09-14:48 UTC (sequences 125-131). 
The second prominence is used as a case study to test the method of cross-correlation of the \ion{Mg}{ii} profiles by \citet{peatprep}.

In this study we focused on the prominence observed on the 28 September 2019. This prominence is at high latitude and corresponds to a very weak filament containing only a few dark features according to observations a few days before in the Meudon spectroheliograms in H$\alpha$ (Figure~\ref{GONG}).
\subsection{SDO}
To investigate the temporal evolution of the prominence at different temperatures corresponding to the chromosphere, transition region and the solar corona, we used the imaging data obtained by AIA in the 304~\AA{}\ (0.1~MK), 171~\AA{}\ (0.6~MK) and 193~\AA{}\ (1.2~MK) channels.
The data presents the full solar disc with a spatial scale of 0.6"/pixel (resolution 1.2"/pixel) at a cadence of 12s.
The pre-processed SDO data were downloaded via JSOC.


\begin{table}[]
\centering
\caption{Characteristics of the IRIS observations.}
\label{IRIS}
\begin{tabular}{@{}lll@{}}
\hline
Pointing & Raster & SJI\\
\hline
x,y = 623'', 777''& FOV 127''x 175'' & FOV 166'' x 175 ''\\
Time= 11:09 - 14:56 & steps 64 x 2'' & \\
 UT & step cadence 16.4 s & SJI cadence 16 s \\
 & Raster Number 13 & SJI Number 832\\
 & Raster cadence 1068s & \\
 \bottomrule 
\label{tab:instr_overview}
\end{tabular}
\end{table}

\section{Morphology}\label{sec:morphology}
\subsection{Cavity}
 We followed the evolution of the prominence using the SJI 2796~\AA{} movie (SJI2796.mp4), and the AIA movies in three different wavelength channels, 171~\AA{} (AIA171.mp4), 193~\AA{} (AIA193.mp4), and 304~\AA{} (AIA304.mp4), after a co-alignement of the images. Snapshots of the movies are presented in Figure~\ref{SJI} and in Appendix~\ref{ap:ap1} (Figures~\ref{AIA_304}, \ref{AIA_171}, \ref{AIA_193}). The SJI 2796~\AA{} movie in particular shows the tremendous dynamics of the fine structures in the prominence. These fine structures are very transient. In all the wavelengths, two horns appear at the top of the prominence between 11:19~UTC and 12:29~UTC. They are no longer visible by 13:15~UTC. The lifetime of these horns is around 90 minutes -- this is consistent with a past study by \citet{Schmit2013a}. They correspond to the bottom of the cavity detectable in X-ray seen by XRT (Figure~\ref{XRT}).

\subsection{IRIS raster}
We reconstructed 13 images from the rasters and computed Dopplershifts and the FWHM assuming that the line profiles are of a Gaussian shape (Figures~\ref{raster_mg2_int} and \ref{raster_mg2_vel}). The horns are visible until 12:53~UTC in the SJI 2796~\AA{} movie. In the raster data, at the beginning of the sequence one horn is blue shifted and the other is redshifted -- suggesting a possible rotational motion. This indicates the presence of a cylindrical flux rope in the cavity visible from its section in the disc plan. Large areas in the prominences are blueshifted and redshifted. This coherence in large areas has already been noted for other prominences \citep{Schmieder2010}.
We produced histograms of Dopplershifts and FWHM. The recovered velocities are mostly around $\pm 20$~km/s (Figure~\ref{fig:iris_histo}). The FWHM values are in the range of $0.2-0.4$~\AA. Few points have a FWHM outside of this range; they are located between the two horns at the bottom of the flux rope.

\begin{figure*}[!htb]
\centering
\includegraphics[width=18cm]{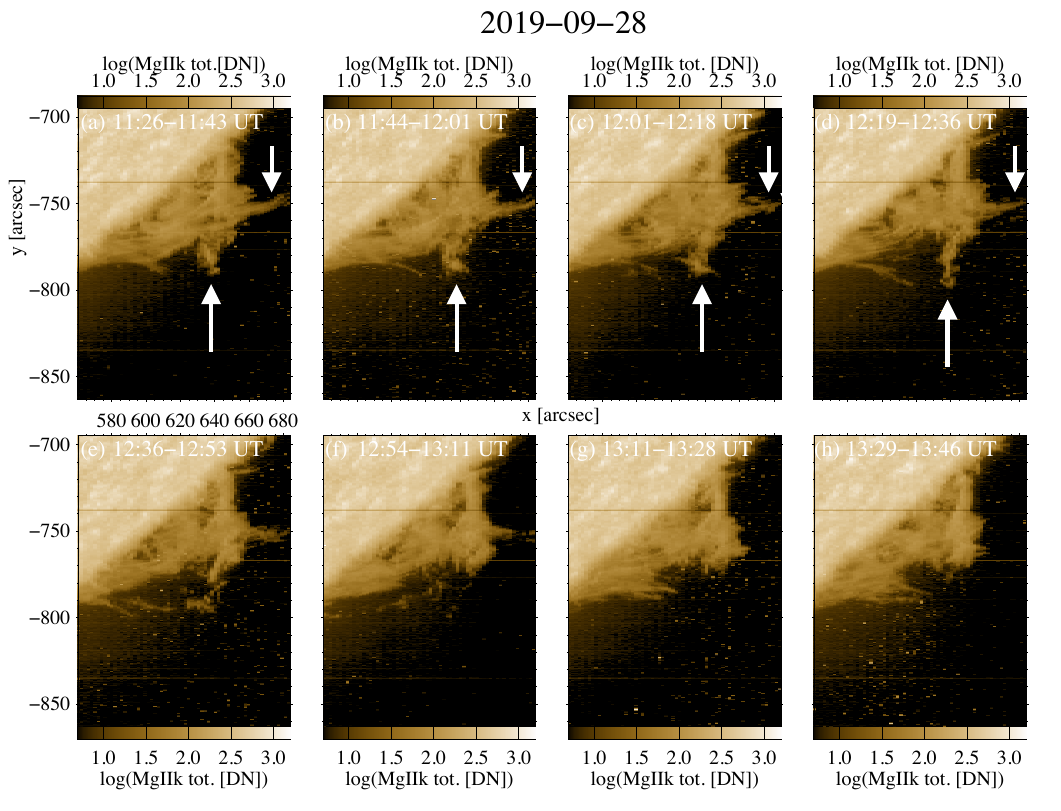}
\caption{Evolution of the \ion{Mg}{ii} k integrated intensity maps from 11:26 UT to 13:46 UT (rasters 1- 8). Each map corresponds to a raster of the prominence obtained in the field of view (127 arc sec x 175 arc sec) indicated in Figure~\ref{SJI} by two vertical dashed lines. The time needed to raster the prominence is indicated at the top of the image (around 16.5 minutes). The maps are reconstructed by using the 64 \ion{Mg}{ii} k slit spectra of the raster scan and integrating the intensity in a wavelength range of $\pm$ 2\AA~around the \ion{Mg}{ii} peak intensity. Two horns are visible at the top of the prominence in the first five maps; white arrows point to the horn's location.}
\label{raster_mg2_int}
\end{figure*}

\begin{figure*}[!htb]
\centering
\includegraphics[width=18cm]{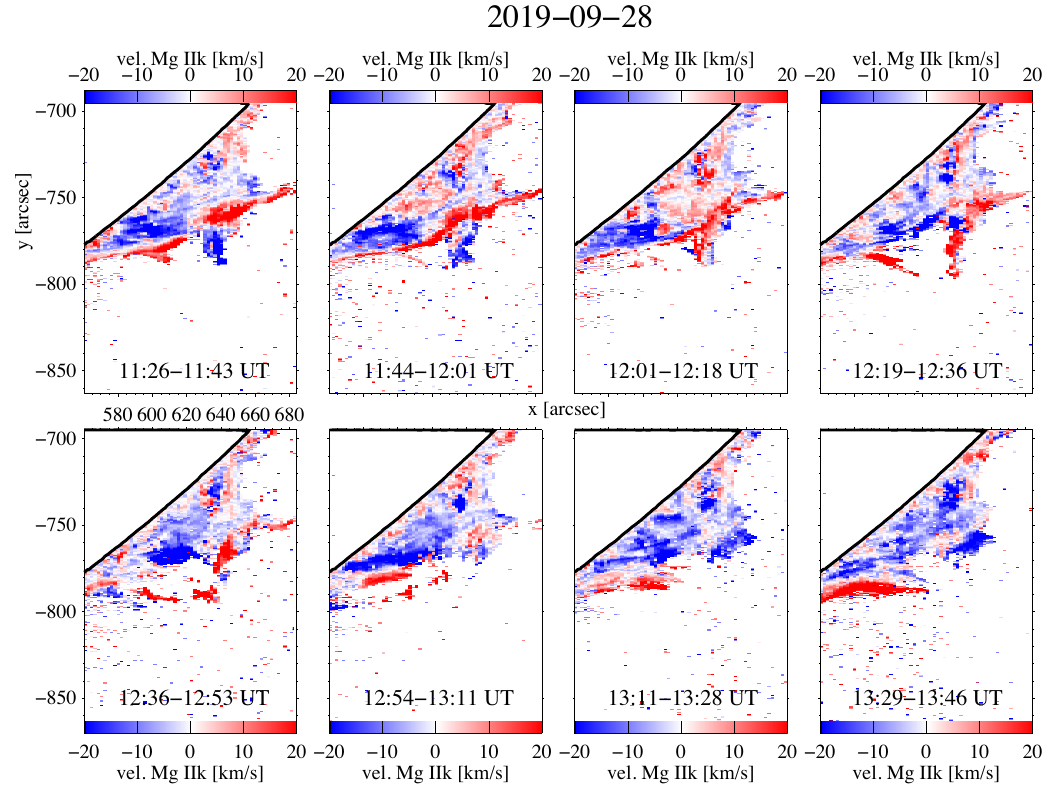}
\caption{Evolution of the \ion{Mg}{ii} k Doppler shift velocity maps from 11:26~UTC to 13:46~UTC (rasters 1- 8). Each map corresponds to a raster of the prominence obtained in the field of view (127 arc sec x 175 arc sec) indicated in Figure~\ref{SJI} by two vertical dashed lines. The Dopplershift velocities are obtained by Gaussian fitting of the profiles.}
\label{raster_mg2_vel}
\end{figure*}

\begin{figure*}[!htb]
\centering
\includegraphics[width=8.8cm]{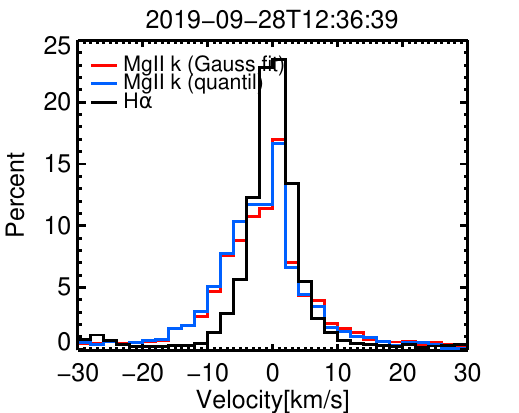}
\caption{Histograms of Doppler velocities for the prominence in H$\alpha$ and in \ion{Mg}{ii} k for the whole \ion{Mg}{ii} prominence using two methods for the computation of Dopplershifts: the Gaussian fit or the quantile method. \label{fig:iris_histo}}
\end{figure*}

\begin{figure*}[!htb]
\centering
\includegraphics[width=\textwidth]{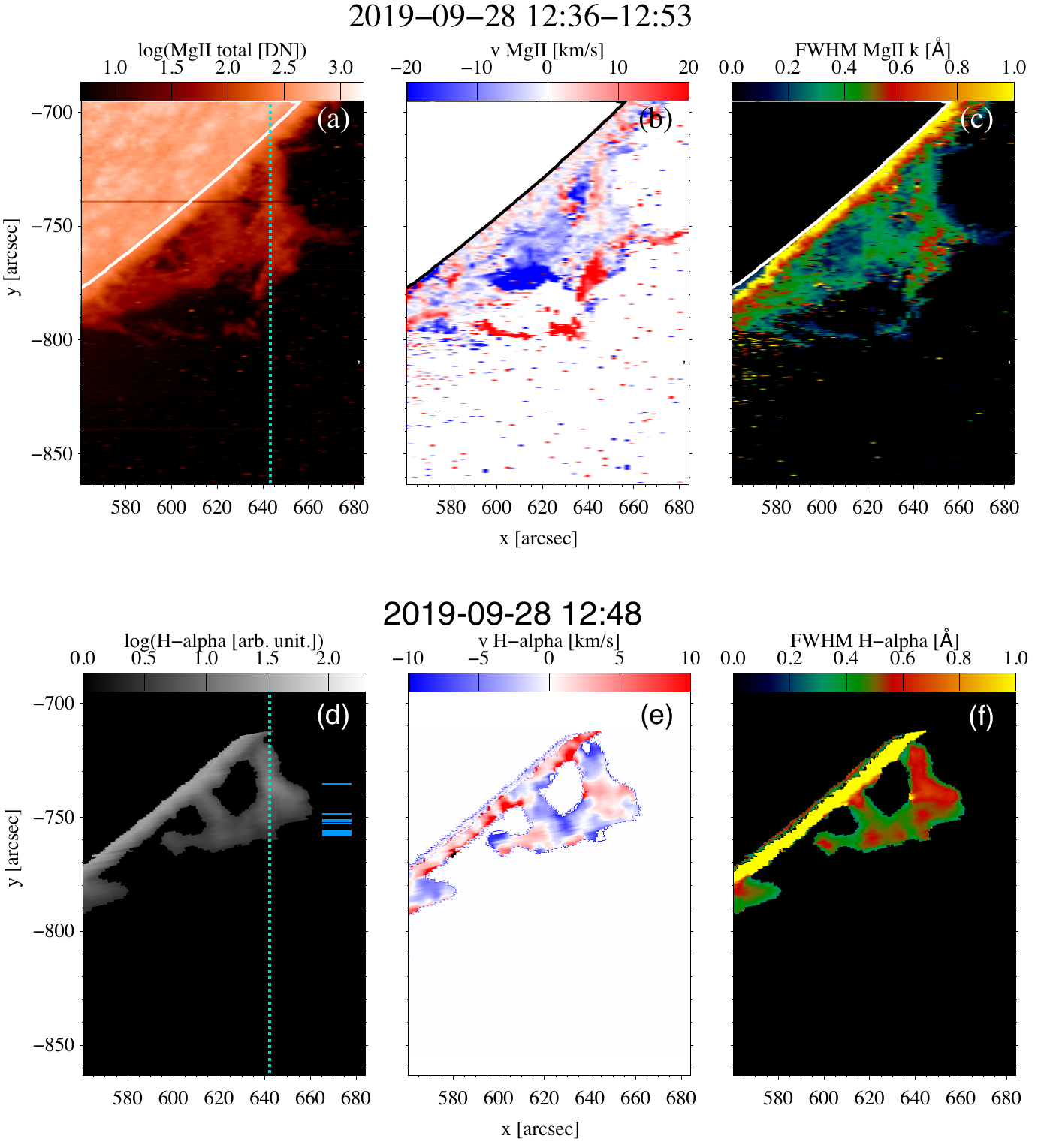}
\caption{Prominence intensity (a), Dopplershift (b), and FWHM (c) images: (top panels) reconstructed from the 64 \ion{Mg}{ii} k spectra of raster 5 obtained by IRIS between 12:36 UT and 12:53 UT; Prominence intensity (d), Dopplershift (e), and FWHM (f) images: (bottom panels) reconstructed from the H$\alpha$ spectra obtained by
THEMIS around 12:48 UT in two minutes. In panel (d) the vertical dashed cyan line corresponds to the slit position 42 in the IRIS raster 5 shown above. The horizontal small blue lines indicate the pixels, B, where the profiles have been analysed (Table 4). The pixels A are not visible in H$\alpha$ image.
\label{themis_iris}}
\end{figure*}

\subsection{Coalignment of IRIS and THEMIS}
We co-aligned the \ion{Mg}{ii} raster data with that of the SJI 2796~\AA{} data. THEMIS H$\alpha$ observations were co-aligned {approximately co-temporally} with AIA 304~\AA,{}\ allowing us to finally co-align {THEMIS} with {the} IRIS \ion{Mg}{ii} SJI.
We used AIA~304 \AA{} to obtain larger context view than provided by IRIS and THEMIS. It was necessary because header information in THEMIS data does not contain pointing information.

To define the limb of the solar disc in IRIS raster data, we computed the total photospheric emission at 2800~\AA\ in a 2~\AA\ large passband and provide a map of such computed total intensity.
Based on this map, we define a region out of disc and out of prominence $[x,y]=[600:640, -810:-840]$~arcsec in raster map obtained between 11:26 and 11:43 UT, hereafter named the background region.
We defined that the solar limb at the location where the photospheric intensity is three times larger than the intensity of the background region.
A so defined solar limb  is presented in Figures~\ref{raster_mg2_vel} and~\ref{themis_iris}a-c.

In the second step, we define the solar limb and the edge of the solar prominence in H$\alpha$.
We compute the total photospheric emission at 6564.38~\AA\ in a 0.6~\AA\ large passband and provide a map of such computed integrated intensity.
Based on this map, we define a region out of disc and out of prominence ($[x,y]=[600:640, -810:-840]$~arcsec in raster map, hereafter named the THEMIS background region).
We define the solar limb as having an intensity two times larger than the mean intensity of the THEMIS background region.%
{ A visual inspection of the photospheric intensity was used to create the definition for the limb used in this paper. However, another method, such as the inflection point method, could be used. Our definition of the limb is consistent with that found by the inflection point method.}
The prominence is defined as the region outside of the limb, and additionally the mean intensity between 6561.85~\AA\ and 6564.07~\AA\ is larger than three times the standard deviation of the THEMIS background region.

{We verified that our method to define the limb  gives identical results to the infletion point method within an error of 1-2 pixels for \ion{Mg}{ii} maps. The THEMIS observations are obtained with a slit scanning the prominence. Hence, we estimate an error bar of a few pixels is in the error bar of the definition of the limb.  The limb is corrugated due to the technical process and seeing; therefore, it is impossible to have a better level of accuracy with the intensity inflection point method than with our threshold method (See Appendix~\ref{ap:ap4}).}

Figure~\ref{themis_iris}
shows the comparison of the prominence in both lines (\ion{Mg}{ii} and H$\alpha$). The horns are observed in \ion{Mg}{ii}, but they are not seen in H$\alpha$ as the material in the horns is too thin to be visible.
In H$\alpha,$ the prominence consists of three columns overlaid by arches. The Dopplershifts in H$\alpha$ are a factor of two lower (Figure~\ref{fig:iris_histo}) than in \ion{Mg}{ii}. This is due to the difference of spatial resolution of the two instruments. Moreover, the structures observed in \ion{Mg}{ii} are different from that observed in H$\alpha$ -- due to their difference in optical depth. H$\alpha$ is generally optically thin in prominences, and therefore the radiation we see is integrated along the line of sight. \ion{Mg}{ii}~h\&k, on the other hand, is typically optically thick, only allowing us to see the front-most material. The FWHM pattern of the prominence in both lines is comparable.

\begin{figure*}[!htb]
\centering
\includegraphics[width=\textwidth]{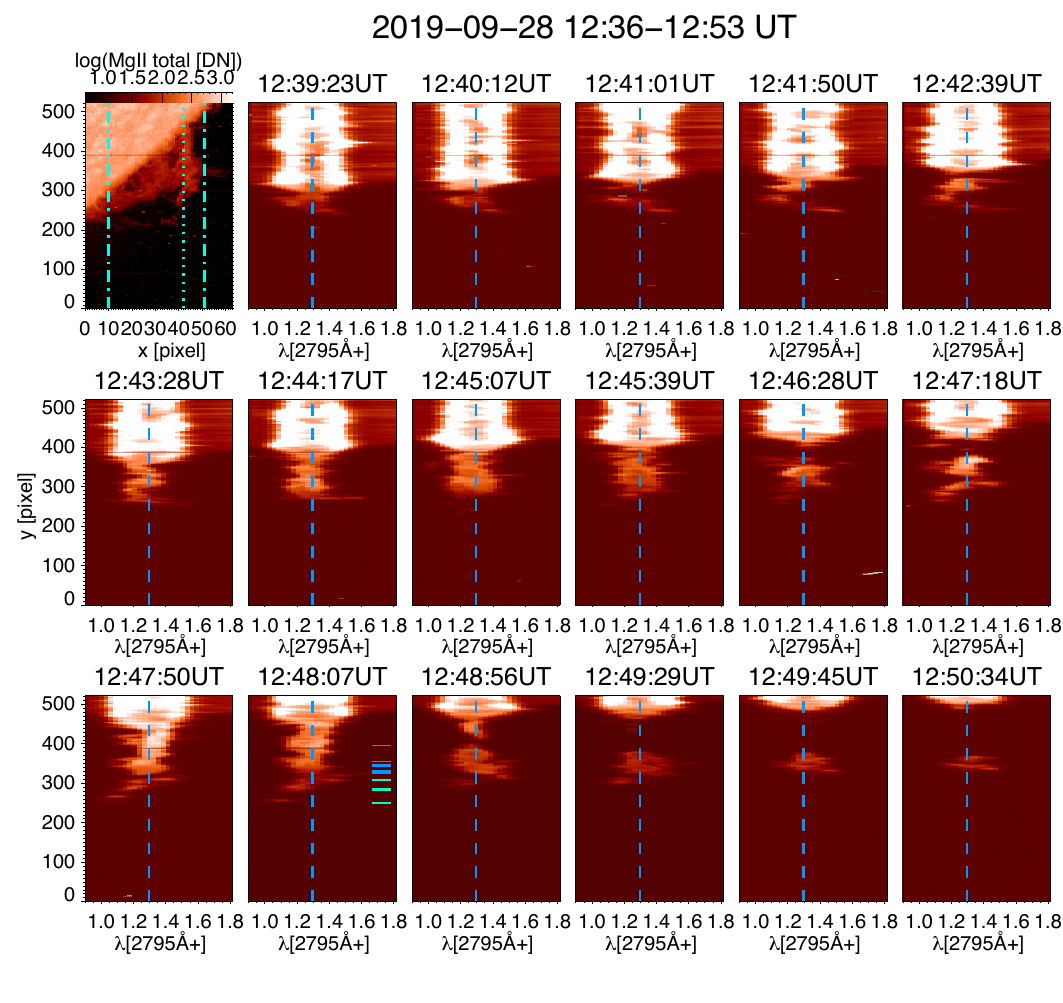}
\caption{Examples of 17 \ion{Mg}{ii} k slit spectra in the scan of raster 5 covering the prominence. The top left image is a slit jaw image, the vertical dashed-dotted lines indicate the FOV of the raster, and the thin dashed line the position of the slit S42 at 12:48~UTC used to draw the profiles in Appendix~\ref{ap:ap3}. In Figures~\ref{fig:IRIS_A} and \ref{fig:IRIS_B} and in the bottom spectra, a series of green and blue lines indicate the position of these profiles along the slit for positions A1-12 and B1-16, respectively.}
\label{fig:IRIS_spectra}
\end{figure*}

\section{Characteristics of H$\alpha$ and \ion{Mg}{ii} profiles}\label{sec:char_ha_mg2}

\subsection{H$\alpha$ profiles}
We used the THEMIS observations obtained at 12:48~UTC
composed by a scan of 512 by 100 pixels.
The scan is performed in the y direction with a step size of 1\arcsec; x is the length of the slit oriented approximately east to west. The length of{ the slit is 
128\arcsec} and corresponds to 512 pixels obtained by oversampling the data by a factor of four. 
From this, 100 H$\alpha$ spectra are observed. 
We calibrated these profiles in wavelength by using the telluric lines and by comparing the intensity of the H$\alpha$ line profile references obtained on the disc. These reference profiles are taken along circles parallel to the limb at $\mu$=0.95, 0.98; 0.99 with the David profiles as a reference, taking into account the $\mu$ value \citep{David1961} (Figure~\ref{fig:calibration} in Appendix~\ref{ap:ap2}). This allows us to compute a proportional factor to multiply the DN number of the{ observations to retrieve the H$\alpha$ profile intensities 
in } erg cm$^{-2}$ sr$^{-1}$ s$^{-1}$ Hz$^{-1}$ (see Appendix~\ref{ap:ap2}).
Some example H$\alpha$ profiles are presented in Appendix~\ref{ap:ap3} (Figure~\ref{Ha_B}).
We reconstruct the THEMIS scan using the integrated intensity and rotate it to north-south to better compare to the IRIS images (Figure~\ref{themis_iris}).

\subsection{Mg\ion{}{ii}~h\&k profiles}
Each IRIS raster consists of 64 slit spectra of \ion{Mg}{ii}~h\&k obtained with a spatial step of 2\arcsec covering 128\arcsec. As an example of this IRIS spectra, Figure~\ref{fig:IRIS_spectra} presents 17 of the 64 slit spectra of raster 5. These presented slit spectra cross the prominence horns. 
After processing the data we obtain radiometrically calibrated \ion{Mg}{ii} profiles.
In the spectra of 12:48:07~UTC in Figure~\ref{fig:IRIS_spectra} -- corresponding to position 42 of the slit -- a handful of the 1D spectra are presented in Appendix~\ref{ap:ap3}.
We were able to match a few IRIS pixels with their corresponding H$\alpha$ pixels. In Appendix~\ref{ap:ap3}, Figure~\ref{fig:IRIS_B} shows the 
\ion{Mg}{ii} k line profiles 
 corresponding to the H$\alpha$ profiles of Figure~\ref{Ha_B}.

\section{Theoretical grid of models}\label{sec:grid}


 {To understand the thermodynamic properties of the prominence, we use the 1D NLTE radiative transfer code, PROM \citep{Gouutebroze1993, heinzel1994}. } {Here, we used the version of PROM presented in \cite{Levens2019}, which builds on the version from \cite{labrosse2004}, which introduced the PCTR. }
 \begin{table}[h]
\centering
\begin{tabular}{ccc} \hline\hline
Parameter & Unit & Value          \\
\hline
$T_{\text{cen}}$ & kK & \begin{tabular}[c]{@{}c@{}}6, 8, 10, 12, 15\\ 20, 25, 35, 40\end{tabular} \\
$T_{\text{tr}}$ & kK & 100         \\
$p_{\text{cen}}$ & dyne cm$^{-2}$ & \begin{tabular}[c]{@{}c@{}}0.01, 0.02, 0.05\\ 0.1, 0.2, 0.5, 1 \end{tabular}\\
        
$p_{\text{tr}}$ & dyne cm$^{-2}$ & 0.01          \\
Slab Width & km & 45 -- 124~100         \\
M & g cm$^{-2}$ & 3.7$\times10^{-8}$ -- 5.1$\times10^{-4}$        \\
H & Mm & 10, 30, 50         \\
$v_{\text{T}}$ & km s$^{-1}$ & 5, 8, 13          \\
$v_{\text{rad}}$ & km s$^{-1}$ & \begin{tabular}[c]{@{}c@{}}0, 2, 4, 6, 8, 10, 20\\40, 60, 80, 100, 150, 200 \end{tabular}\\

$\gamma$ & & 0, 2, 4, 5, 10 \\ \hline \hline        
\end{tabular}
\caption{Model parameters. We note that not all of these are uniquely combined.}
\label{Table_parameters}
\end{table}

 {We used a grid of 23~940 models that produce \ion{Mg}{ii}~h\&k and \ion{H}{i} spectra \citep{peatprep}. The models comprise both isothermal and isobaric atmospheres and atmospheres with a prominence-to-corona transition region (PCTR). This large grid of models allows us to explore a larger domain of parameters than in previous studies \citep{Peat2021, Barczynski2021}. The parameters of these models can be seen in Table \ref{Table_parameters}. $T_{\text{cen}}$ and $p_{\text{cen}}$ are the central temperature and pressure, respectively; $T_{\text{tr}}$ and $p_{\text{tr}}$ are the temperature and pressure at the edge of the PCTR, respectively; slab width is the width of the slab; $M$ is the column mass; $H$ is the height above the solar surface; $v_T$ is the microturbulent velocity; $v_{\text{rad}}$ is the outwards radial velocity of the slab; and $\gamma$ is a dimensionless number that dictates the extent of the PCTR. A $\gamma$ value of zero indicates the model is isothermal and isobaric. For isothermal and isobaric models, $T_{\text{tr}}\equiv T_{\text{cen}}$, and $p_{\text{tr}}\equiv p_{\text{cen}}$.}

 {The version of PROM includes a PCTR that is formulated as a function of column mass as in \cite{Anzer1999}:}
\begin{equation} 
 T(m)=T_{\text{cen}}+(T_{\text{tr}}-T_{\text{cen}})\left(1-4\frac{m}{M}\left(1-\frac{m}{M}\right)\right)^\gamma,
 \label{tstrat}
\end{equation}
\begin{equation}
 p(m)=4p_c\frac{m}{M}\left(1-\frac{m}{M}\right)+p_{\text{tr}},
 \label{pstrat}
\end{equation}
where $p_c=p_{\text{cen}}-p_{\text{tr}}$ and $\gamma\geq2$. {These equations are not used for the isobaric and isothermal atmospheres, when} $\gamma=0$ {it is} simply a placeholder indicating that the model is isobaric and isothermal. The PCTR pressure profile is derived from magnetohydrostatic equilibrium, whereas the temperature profile is empirical.
Additionally, the \ion{Ca}{ii} atom from \cite{gouttebroze_1997} was used as the foundation on which \cite{Levens2019} built the \ion{Mg}{ii} ion. The prominence is illuminated on both sides by the solar disc; this incident radiation is taken from disc centre IRIS observations on 29 September 2013 \citep{Levens2019}. 
All the details of the used code PROM are presented in \citet{Levens2019}.

A parametric study of the variation of the parameters in  the space defined by  Table 3  is suitable but  outside the scope of this study. Such a study is planned for \cite{peatprep}.

\subsection{Theoretical relationship between H$\alpha$ and \ion{Mg}{ii} intensities}

The new grid of 23~940 models expand on and include the original 1007 models from \citet{Peat2021}. These models allow us to compute the relationship between the integrated \ion{Mg}{ii} intensities (E(\ion{Mg}{ii})) and the integrated H$\alpha$ intensities (E(H$\alpha$)) and segregate the models according to mean temperature and thickness.
Figure~\ref{EM} (left panel) shows this relationship between the computed values of E(\ion{Mg}{ii}k) and E(H$\alpha$) for a microturbulent velocity of 5 km~s$^{-1}$. The relationship between E(\ion{Mg}{ii}k) and the emission measure (EM) is presented in the top right panel, and the relationship between E(H$\alpha$) and EM is presented in the bottom right panel. { EM is defined by
\begin{equation}
 EM=\int^D_0 {n_e}^2dz \ .
 \label{eq:em}
\end{equation}
}
For comparison with the observations, we selected pixels belonging to the fine structure of the prominence.

\subsection{Comparing observations and synthetic profiles}

\citet{peatprep} worked to improve the rRMS method presented in \citet{Peat2021}. The `rolling' aspect of the algorithm was replaced with a cross-correlation to predetermine the `roll'. This increased the computation speed by a factor of ten. Due to the removal of the rolling aspect, the procedure was renamed cross-RMS (xRMS) to account for the new cross-correlation. 
We obtain a relatively 
good match between observed and synthesised profiles for all the rasters.

As an example, in raster 5, good matches are found for 45~\% of the pixels with an RMS value less than 15~000, 22.5~\% of which are fitted with models with a PCTR of a higher mean temperature. However, it is difficult to find a 1D model that fits the \ion{Mg}{ii} profiles in the main core of the prominence due to its complex profiles, as shown in Figure~\ref{classification}. This is why we concentrated our study of the parameters on the horns and edge of the prominence.

\begin{table*}[]
\caption{Characteristics of \ion{Mg}{ii} k and h line profiles of IRIS raster 5 (slit 42 at 12:48:07 UT) and simultaneous H$\alpha$ line profiles obtained with the THEMIS. { The selected points can be visualised in Figure \ref{classification}.}}
\label{tab:Obs}
\begin{tabular}{@{}lll|cccll|lll@{}}
\hline
\multicolumn{7}{c|}{IRIS (\ion{Mg}{ii})}&\multicolumn{3}{c}{MSDP (H$\alpha$)} \\
\hline
Number & Y& $E$(\ion{Mg}{ii} k) \tablefootmark{a} & $E$(\ion{Mg}{ii} h) \tablefootmark{b} & R(k/h) \tablefootmark{c} &Velocity & FWHM & E(H$\alpha$) \tablefootmark{d} & Velocity & FWHM \\
 & &{[}px{]} & & & &{[}km~s$^{-1}${]} &~\AA{} & & {[}km~s$^{-1}${]} &~\AA{}\\
 \hline
 A1             &       249     &       0.71    &       0.51    &       1.40    &       -13.86  &       0.21    &       -       &       -       &       -       \\
A2              &       250     &       1.01    &       0.73    &       1.40    &       -13.04  &       0.23    &       -       &       -       &       -       \\
A3              &       251     &       1.18    &       0.82    &       1.44    &       -13.39  &       0.27    &       -       &       -       &       -       \\
A4              &       252     &       1.07    &       0.76    &       1.41    &       -13.39  &       0.28    &       -       &       -       &       -       \\
A5              &       253     &       0.72    &       0.53    &       1.36    &       -16.41  &       0.29    &       -       &       -       &       -       \\
A6              &       283     &       1.40    &       0.92    &       1.53    &       15.14   &       0.39    &       -       &       -       &       -       \\
A7              &       284     &       1.46    &       1.01    &       1.46    &       19.47   &       0.31    &       -       &       -       &       -       \\
A8              &       285     &       1.66    &       1.13    &       1.47    &       20.41   &       0.30    &       -       &       -       &       -       \\
A9              &       286     &       1.95    &       1.29    &       1.51    &       21.28   &       0.36    &       -       &       -       &       -       \\
A10             &       287     &       2.23    &       1.47    &       1.52    &       23.39   &       0.40    &       -       &       -       &       -       \\
A11             &       308     &       1.75    &       1.10    &       1.59    &       28.03   &       0.14    &       -       &       -       &       -       \\
A12             &       309     &       1.61    &       1.03    &       1.56    &       27.95   &       0.15    &       -       &       -       &       -       \\
\hline                                                                  

B1              &       326     &       2.06    &       1.52    &       1.35    &       -4.87   &       0.31    &       2.91    &       3.31    &       0.62    \\
B2              &       327     &       2.05    &       1.52    &       1.35    &       -4.78   &       0.32    &       2.41    &       3.56    &       0.62    \\
B3              &       328     &       2.17    &       1.54    &       1.42    &       -4.73   &       0.33    &       2.08    &       3.34    &       0.64    \\
B4              &       329     &       2.22    &       1.54    &       1.44    &       -3.82   &       0.31    &       2.03    &       2.89    &       0.65    \\
B5              &       330     &       2.29    &       1.62    &       1.42    &       -2.18   &       0.31    &       1.60    &       0.82    &       0.59    \\
B6              &       331     &       2.41    &       1.73    &       1.39    &       -2.45   &       0.32    &       1.49    &       0.60    &       0.53    \\
B7              &       332     &       2.50    &       1.80    &       1.39    &       -2.35   &       0.33    &       1.46    &       0.50    &       0.52    \\
B8              &       333     &       2.61    &       1.91    &       1.37    &       -2.12   &       0.34    &       1.29    &       1.82    &       0.51    \\
B9              &       342     &       3.12    &       2.21    &       1.41    &       -3.08   &       0.35    &       1.37    &       0.27    &       0.57    \\
B10             &       343     &       2.88    &       2.07    &       1.39    &       -0.37   &       0.32    &       1.33    &       -0.16   &       0.56    \\
B11             &       346     &       3.21    &       2.15    &       1.49    &       -4.45   &       0.36    &       1.37    &       -2.18   &       0.55    \\
B12             &       347     &       3.04    &       2.11    &       1.44    &       -2.63   &       0.33    &       1.30    &       -2.27   &       0.55    \\
B13             &       348     &       2.90    &       2.02    &       1.44    &       -1.16   &       0.30    &       1.03    &       -2.12   &       0.53    \\
B14     &       356     &       3.11    &       2.12    &       1.47    &       -0.27   &       0.35    &       0.77    &       -1.71   &       0.52    \\
B15             &       396     &       2.96    &       1.95    &       1.52    &       16.03   &       0.31    &       0.66    &       -2.16   &       0.51    \\
B16             &       397     &       2.90    &       1.86    &       1.56    &       16.19   &       0.33    &       0.45    &       -6.34   &       0.51    \\
                                                                                 \hline
\end{tabular}
\tablefoot{\\
\tablefoottext{(a)}{Integrated intensity of \ion{Mg}{ii} k line between 2795.49~\AA\ and 2797.53~\AA\ in erg sr$^{-1}$ s$^{-1}$ cm$^{-2}$ $\times$ 10$^{4}$}\\
\tablefoottext{(b)}{Integrated intensity of \ion{Mg}{ii} h line between 2802.57\AA\ and 2804.61~\AA\ in erg sr$^{-1}$ s$^{-1}$ cm$^{-2}$ $\times$ 10$^{4}$}\\
\tablefoottext{(c)}{Ratio between $E$(\ion{Mg}{ii} k) and $E$(\ion{Mg}{ii} h)}\\
\tablefoottext{(d)}{Integrated intensity of H$\alpha$ line in $10^{4}$ erg sr$^{-1}$ s$^{-1}$ cm$^{-2}$.}
}
\end{table*}

\begin{table*}[ht!]
\caption{Plasma parameters of the selected prominence and horn pixels in raster 5 and slit at 12:48:07 UT (slit 42 counting from 0) from non-LTE radiative transfer models. Fits with an RMS higher than 15~000 are considered unsatisfactory. { The selected pixels are visualized in Figure \ref{classification}. The columns labelled $\tau$(h) and $\tau$(k) give the optical thickness in \ion{Mg}{ii}~h\&k lines, respectively.}}
\begin{tabular} {lllllrlllll}
\hline
 {Num} & {Pix} & $Ne$ &$\tau$(h)& $\tau$(k)& Mean $T$& Mean $P$ &FWHM h &FWHM k &$\tau$ (H$\alpha$) & RMS \\
& & 10$^{9}$ cm$^{-3}$ & & & K & dyne cm$^{-2}$ &~\AA{} &~\AA{} &10$^{-3}$ & \\
\hline
A1 & 249 & 19.92 & 0.16 & 0.32 & 40909 & 0.22 & 0.1135 & 0.1137 & 92.00 & 4602 \\
A2 & 250 & 13.45 & 2.90 & 5.90 & 8000 & 0.05 & 0.1648 & 0.1847 & 35.00 & 14240 \\
A3 & 251 & 2.93 & 1.60 & 3.10 & 16443 & 0.02 & 0.2308 & 0.2549 & 6.60 & 11155 \\
A4 & 252 & 2.02 & 1.20 & 2.40 & 21777 & 0.01 & 0.2360 & 0.2551 & 2.30 & 19024 \\
A5 & 253 & 5.40 & 0.56 & 1.10 & 12000 & 0.02 & 0.1459 & 0.1521 & 2.50 & 8939 \\
A6 & 283 & 5.49 & 0.35 & 0.70 & 12000 & 0.02 & 0.2209 & 0.2277 & 2.10 & 13187 \\
A7 & 284 & 6.69 & 1.00 & 2.00 & 8000 & 0.02 & 0.1453 & 0.1565 & 7.50 & 9590 \\
A8 & 285 & 7.81 & 1.50 & 3.00 & 6000 & 0.02 & 0.1494 & 0.1638 & 16.00 & 18075 \\
A9 & 286 & 2.30 & 1.50 & 3.10 & 16443 & 0.02 & 0.2296 & 0.2536 & 4.60 & 19218 \\
A10 & 287 & 13.51 & 1.90 & 3.70 & 8000 & 0.05 & 0.2354 & 0.2619 & 27.00 & 24221 \\
A11 & 308 & 13.22 & 0.93 & 1.80 & 27269 & 0.10 & 0.1067 & 0.1132 & 35.00 & 20564 \\
A12 & 309 & 22.93 & 2.50 & 5.00 & 15000 & 0.10 & 0.1139 & 0.1269 & 33.00 & 15995 \\ \hline
B1 & 326 & 13.94 & 22.00 & 44.00 & 6000 & 0.05 & 0.2172 & 0.2357 & 320.00 & 16443 \\
B2 & 327 & 14.73 & 2.70 & 5.40 & 6000 & 0.05 & 0.2482 & 0.2785 & 50.00 & 15184 \\
B3 & 328 & 14.73 & 2.70 & 5.40 & 6000 & 0.05 & 0.2482 & 0.2785 & 50.00 & 16611 \\
B4 & 329 & 2.10 & 1.10 & 2.30 & 24801 & 0.02 & 0.2233 & 0.2433 & 3.90 & 15619 \\
B5 & 330 & 2.16 & 18.00 & 36.00 & 24801 & 0.02 & 0.2141 & 0.2332 & 77.00 & 15083 \\
B6 & 331 & 5.95 & 2.40 & 4.80 & 10000 & 0.02 & 0.2452 & 0.2745 & 19.00 & 15390 \\
B7 & 332 & 4.66 & 2.00 & 4.00 & 15000 & 0.02 & 0.2416 & 0.2692 & 13.00 & 11728 \\
B8 & 333 & 23.31 & 210.00 & 420.00 & 6000 & 0.20 & 0.2802 & 0.3007 & 1800.00 & 24531 \\
B9 & 342 & 2.30 & 17.00 & 34.00 & 26401 & 0.02 & 0.2145 & 0.2353 & 64.00 & 24300 \\
B10 & 343 & 6.28 & 2.20 & 4.50 & 10000 & 0.02 & 0.2445 & 0.2731 & 17.00 & 18346 \\
B11 & 346 & 36.02 & 270.00 & 540.00 & 6000 & 0.50 & 0.2928 & 0.3153 & 1900.00 & 28707 \\
B12 & 347 & 23.84 & 92.00 & 180.00 & 6000 & 0.10 & 0.2593 & 0.2780 & 1400.00 & 29146 \\
B13 & 348 & 6.82 & 1.60 & 3.20 & 16443 & 0.06 & 0.2323 & 0.2561 & 19.00 & 15129 \\
B14 & 356 & 11.33 & 1.30 & 2.60 & 21777 & 0.08 & 0.2354 & 0.2547 & 37.00 & 38344 \\
B15 & 396 & 3.74 & 4.70 & 9.30 & 31991 & 0.03 & 0.1276 & 0.1436 & 24.00 & 26640 \\
B16 & 397 & 6.15 & 5.10 & 10.00 & 19027 & 0.03 & 0.1296 & 0.1461 & 20.00 & 19463 \\ \hline
\end{tabular}
\label{tab:mg2_ha_sel}
\end{table*}

\begin{table*}[]
\caption{Complementary plasma parameters of the fits in the selected pixels in the prominence and horns (raster 5, slit 42 at 12:48 UT) from non-LTE radiative transfer models: central temperature, central gas pressure, mean temperature, mean gas pressure, mean electron density, mean N$_H$, $\gamma$. For verification purposes, some parameters of Table \ref{tab:mg2_ha_sel} are again given.}
\centering
\begin{tabular}{llrrrrrrr} \hline\hline
Num & Pix & Cent Temp & Cent Pres & Mean Temp & Mean Pres & Mean n$_\text{e}$ & Mean n$_\text{HI}$ & $\gamma$ \\
 & & K & dyn~cm$^{-2}$ & K & dyn~cm$^{-2}$ & 10$^9$~cm$^{-3}$ & 10$^8$~cm$^{-3}$ & \\\hline
A1 & 249 & 35000.0 & 1.0 & 40908.72 & 0.22 & 1.99 & 0.01 & 5.0 \\
A2 & 250 & 8000.0 & 0.05 & 8000.00 & 0.05 & 1.35 & 15.47 & 0.0 \\
A3 & 251 & 6000.0 & 0.10 & 16443.12 & 0.02 & 0.29 & 2.67 & 4.0 \\
A4 & 252 & 12000.0 & 0.02 & 21776.55 & 0.01 & 0.20 & 0.30 & 4.0 \\
A5 & 253 & 12000.0 & 0.02 & 12000.00 & 0.02 & 0.54 & 1.26 & 0.0 \\
A6 & 283 & 12000.0 & 0.02 & 12000.00 & 0.02 & 0.55 & 1.08 & 0.0 \\
A7 & 284 & 8000.0 & 0.02 & 8000.00 & 0.02 & 0.67 & 3.68 & 0.0 \\
A8 & 285 & 6000.0 & 0.02 & 6000.00 & 0.02 & 0.78 & 7.03 & 0.0 \\
A9 & 286 & 6000.0 & 0.05 & 16443.12 & 0.02 & 0.23 & 1.79 & 4.0 \\
A10 & 287 & 8000.0 & 0.05 & 8000.00 & 0.05 & 1.35 & 15.37 & 0.0 \\
A11 & 308 & 20000.0 & 0.5 & 27269.26 & 0.10 & 1.32 & 0.16 & 5.0 \\
A12 & 309 & 15000.0 & 0.10 & 15000.00 & 0.10 & 2.29 & 2.43 & 0.0 \\\hline
B1 & 326 & 6000.0 & 0.05 & 6000.00 & 0.05 & 1.39 & 28.17 & 0.0 \\
B2 & 327 & 6000.0 & 0.05 & 6000.00 & 0.05 & 1.47 & 26.72 & 0.0 \\
B3 & 328 & 6000.0 & 0.05 & 6000.00 & 0.05 & 1.47 & 26.72 & 0.0 \\
B4 & 329 & 6000.0 & 0.05 & 24800.61 & 0.02 & 0.21 & 0.82 & 2.0 \\
B5 & 330 & 6000.0 & 0.05 & 24800.61 & 0.02 & 0.22 & 0.75 & 2.0 \\
B6 & 331 & 10000.0 & 0.02 & 10000.00 & 0.02 & 0.59 & 2.12 & 0.0 \\
B7 & 332 & 15000.0 & 0.02 & 15000.00 & 0.02 & 0.47 & 0.35 & 0.0 \\
B8 & 333 & 6000.0 & 0.20 & 6000.00 & 0.20 & 2.33 & 174.88 & 0.0 \\
B9 & 342 & 8000.0 & 0.05 & 26400.60 & 0.02 & 0.23 & 0.52 & 2.0 \\
B10 & 343 & 10000.0 & 0.02 & 10000.00 & 0.02 & 0.63 & 1.48 & 0.0 \\
B11 & 346 & 6000.0 & 0.5 & 6000.00 & 0.5 & 3.60 & 479.59 & 0.0 \\
B12& 347 & 6000.0 & 0.10 & 6000.00 & 0.10 & 2.38 & 64.20 & 0.0 \\
B13& 348 & 6000.0 & 0.5 & 16443.12 & 0.06 & 0.68 & 8.43 & 4.0 \\
B14& 356 & 12000.0 & 0.5 & 21776.55 & 0.08 & 1.13 & 2.71 & 4.0 \\
B15 & 396 & 15000.0 & 0.10 & 31990.73 & 0.03 & 0.37 & 0.11 & 2.0 \\
B16 & 397 & 15000.0 & 0.10 & 19027.29 & 0.03 & 0.61 & 0.44 & 10.0 \\\hline
\end{tabular}
\label{tab:parameters}
\end{table*}



\subsection{Horn and edge profile characteristics}
As mentioned previously, we worked on a select few IRIS \ion{Mg}{ii}~h\&k profiles along slit 42 in raster 5, which corresponds to a time of 12:48:07 UT.
Slit 42 crosses the column edge and one of the horns of the prominence.
The selection of the \ion{Mg}{ii}~h\&k profiles is based on three criteria: the peak intensity should be not lower than 5 {$ \times$ $10^{4}$ erg sr$^{-1}$ s$^{-1}$ \AA$^{-1}$ cm$^{-2}$}; the \ion{Mg}{ii}~h\&k profile should have a main single peak; the fitted model profile must not exceed a main temperature of 30~000 K.

In Table~\ref{tab:Obs} we present our selection. The A profiles are located in the horn (between 249 and 309 y-pixel positions). The B profiles are in the column or edge of the column of the prominence (between 326 and 397 y-pixel positions).
Figure~\ref{r5396} shows two profiles: one (A2) in the horn, which is well fitted by one synthetic profile and the RMS value is low; and another (B11) in the prominence body which is not well fitted, with a large RMS value.
This shows how the model chosen by the code is very sensitive to the shape of the profile. 
The RMS values are around 4000 to 20000 for the horn and much larger than this in the main body of the prominence and even at the edge of the prominence (pixels B1-16). We observe that the peak intensity of the horn profiles is much lower than the prominence by a factor of two. 

 All the pixels (A1-12 and B1-16) of the selection did not completely match with our three criteria. For example, for two pixels (A1, B15) the mean temperature of the models exceeds 30000 K.
 For half the pixels in our selection the RMS value exceeds 15000 but we keep them because it is nearly impossible to fit these profiles with a single Gaussian (see pixel B 11 in Figure~\ref{r5396}).
 %

For the A profiles, no H$\alpha$ emission was identified in the THEMIS observations.
The H$\alpha$ prominence is limited to areas where the prominence is geometrically thick enough.
This has been previously found in earlier studies \citep{heinzel2001b,Barczynski2021}. 
A large area extending out of the H$\alpha$ prominence contour is detected in the \ion{Mg}{ii}~h\&k lines, this corresponds to the horns. 
The characteristics of the horns were analysed using only the \ion{Mg}{ii}~h\&k line profiles (Section 6.2).
Finally, our selection for comparison of H$\alpha$ and \ion{Mg}{ii}~h\&k is reduced to the fine structure seen in H$\alpha$ where \ion{Mg}{ii}~h\&k profiles exhibit mostly single peak (B profiles).
These pixels correspond relatively to thin single threads.

\subsection{Results of non-LTE modelling}
 {When discussing temperatures and pressures here, we are strictly speaking of the mean temperature and pressure -- defined by \citep{Peat2023},
\begin{equation}
 \overline{p}=\frac{2p_{\text{cen}}+p_{\text{tr}}}{3},
 \label{meanpres}
\end{equation}
and
\begin{equation}
 \overline{T}=\frac{2\gamma T_{\text{cen}}+T_{\text{tr}}}{2\gamma+1}.
 \label{meantemp}
\end{equation}
For PCTR models, it should be noted that this is a mean temperature, that is parts of the prominence atmosphere exist at lower and higher temperatures. }
The results found for each point after choosing the best models for fitting \ion{Mg}{ii} profiles are presented in Table \ref{tab:mg2_ha_sel}. 
The values of $\tau$(k) are very small between points 1 and 5 when the profiles are fitted with models with PCTR and high temperatures. On the other hand, the values of $\tau$(k) may reach 40 to 500 when the profiles are fitted with isobaric models.
There are significant variations of solutions for adjacent pixels
{ due to the mode of IRIS observations with a large x step (2").}
We find relatively high values of $\tau$(k) for pixels fitted with isobaric models in comparison to the prominence tornado, which used the rRMS technique \citep{Peat2021} where the highest value of $\tau$(k) are around 60.

\section{Physical parameters}\label{sec:phys_par}
\subsection{Emission measure from H$\alpha$ and \ion{Mg}{ii}}

The observed integrated intensities of the \ion{Mg}{ii}~k line and H$\alpha$ can be compared with 
 the general trend obtained with the theoretical models (Figure~\ref{EM}). 
 As we explain in Section 5.4, the A profiles have no counterpart in H$\alpha$; they are located in the horns. The B profiles, however, do have counterparts in H$\alpha$ and are located at the edge of the prominence.
 In this study, we did not consider the main columns of the prominence visible in H$\alpha$ and \ion{Mg}{ii} because the fitting of the profiles is nearly impossible for profiles with many peaks instead of one single peak. A multithread model would be necessary to have a good fitting, but this is very computationally expensive. Table~\ref{tab:mg2_ha_sel} presents pixels with a mix of good and 'bad' fits in \ion{Mg}{ii}~h\&k profiles.
 { However, the selected B points still have a reasonable rms (only twice the threshold) that we consider acceptable (Figure \ref{r5396} point B11).\\
 
 Figure~\ref{EM} shows models with smaller 
Mg II integrated intensities compared with graphs of the previous study of \citet{Barczynski2021}.
In that figure 
we draw the limits of the integrated intensity values for the B pixels using the \ion{Mg}{ii}~k and H$\alpha$ measurements from Table \ref{tab:Obs}. 
The models with small thread diameters of 0$-$1000 km and temperature range ({6000 K}- 20000K) are found to be consistent with our observational bounds.
The emission measure values found from the right hand panels are in the range of $7\times 10^{27}<EM<1.5 \times 10^{29}$~cm$^{-5}$ corresponding to E(Mg II k) and $4 \times 10^{27}<EM< 8 \times 10^{28}$~cm$^{-5}$ for E(H$\alpha$). 

This corresponds to a relatively large range of electron densities between $5 \times 10^{9}$ and $5 \times 10^{10}$cm$^{-3}$.}
This uncertainty is rather large.

\subsection{Plasma parameters from fitting \ion{Mg}{ii} profiles} 
 { In Tables \ref{tab:mg2_ha_sel} and \ref{tab:parameters}} we present the relevant outputs of the modelling for the selected pixels in the horns (A) and
{ at the edge of the prominence} (B) where the prominence is visible in H$\alpha$.
We note that the threshold of the xRMS method is reached in both cases for some pixels; in zone A, seven (out of twelve) have a fitting with a rms lower than 15000, and in zone B nearly all have a rms larger than 15000. This means that it is difficult to fit the observed profiles in B, mainly because they have two peaks and a self reversal as in Figure \ref{r5396} (B11). Therefore, the values of the parameters of pixels B are only given as an estimation or a trend value. 
The FWHM (k) reaches 0.3~\AA\ in many pixels of zone B contrary to pixels where the FWHM (k) is between 0.11 and 0.25~\AA\ with single peak profiles. { This suggests that the line-of-sight crosses several fine structures}.

We note that more isobaric, isothermal models are found to fit \ion{Mg}{ii} profiles in zone B (six out of ten). The mean temperature is between 6000 and 8000~K. The optical thickness of \ion{the Mg}{ii} k line can be large (540 for B11). The corresponding optical thickness for H$\alpha$ is around one, which is consistent with the observations, and the prominence is visible in H$\alpha$ in zone B. \ion{Mg}{ii} lines are optically thicker, and along the LOS only a part of the prominence is observed. 
The electron density in zone A is around $2\times 10^{9}$ cm$^{-3}$ with a few points of $1.3-2\times 10^{10}$~cm$^{-3}$, while for zone B it is around $2\times 10^{10}$ cm$^{-3}$ and a few values of $5\times 10^{9}$ cm$^{-3}$. 

 {The conclusions obtained from the outputs of the modelling should be taken with caution. For example, the incident radiation used in our non-LTE calculations may not accurately represent the true incident radiation on the observed prominence. This may have an effect on the results as discussed by \citet{Gunar2022}. There are also other uncertainties associated with IRIS radiometric calibration and degradation of the detector over time. The radiometric calibration is determined using spectral radiance measurements from the International Ultraviolet Explorer \citep[IUE;][]{iue1978} -- the radiances of which are quoted as having an uncertainty of 10-15\% \citep{tian_itn_2014}.}

\subsection{Comparison with previous studies}
 {Our EM values are { in a first approximation} similar to those found by other authors \citep{Gouutebroze1993,heinzel1994,Jejcic2018}. However, we note that our range of EM values is narrower.} 

Using the fitting profile method { (Tables \ref{tab:mg2_ha_sel} and \ref{tab:parameters})}, in zone B we find values of electron density ($\sim 10^{10}$ cm$^{-3}$) typical of prominences when the prominence is also visible in H$\alpha$ using the classical method based on the emission measure, as presented in the previous paragraph or in other studies \citep{Ruan2019,Jejcic2022}.
In zone A, we see a similar behaviour of the physical parameters found in the top of a tornado prominence with lower electron density except in a few points with value exceeding 
$2\times 10^{10}$ cm$^{-3}$ \citep{Barczynski2021}. However, it is found for 
 narrower threads. 

Our method of fitting full profiles succeeds to reach such single threads which have very specific physical characteristics. These profiles correspond to thin threads with PCTRs that have an optical thickness in H$\alpha$ too low to be detectable and may be too hot to be visible in H$\alpha$. 

There are still a few pixels (A10, A11, A12) showing a higher electron density, higher pressure (0.1 dyne cm$^{-2}$), and a higher temperature in the horn.
When the temperature is over 15~000 K, \citet{Peat2021} concluded that the inclusion of a PCTR can dramatically influence the value of the ionisation degree {\citep[n(\ion{H}{ii})/n(\ion{H}{i});][]{th1995}} and electron density. {Past studies \citep[such as][]{vial1998, Zhang2019} have shown that the ionisation degree generally does not rise above ten. However, such studies do not exceed temperatures of 15~000K, where the ionisation degree is seen to exponentially rise \citep{Peat2021}. This was likely due to the isothermal and isobaric nature of the models used, and that \ion{Mg}{ii} is said to ionise to \ion{Mg}{iii} in the 15~000~K to 30~000~K range (see Figure 7 in \citet{Heinzel2014}). This kind of behaviour was found in the tornado and interpreted as{ plasma in the condensation phase} \citep{Barczynski2021}.

\begin{figure*}[!tb] 
\center{\includegraphics[width=18cm]{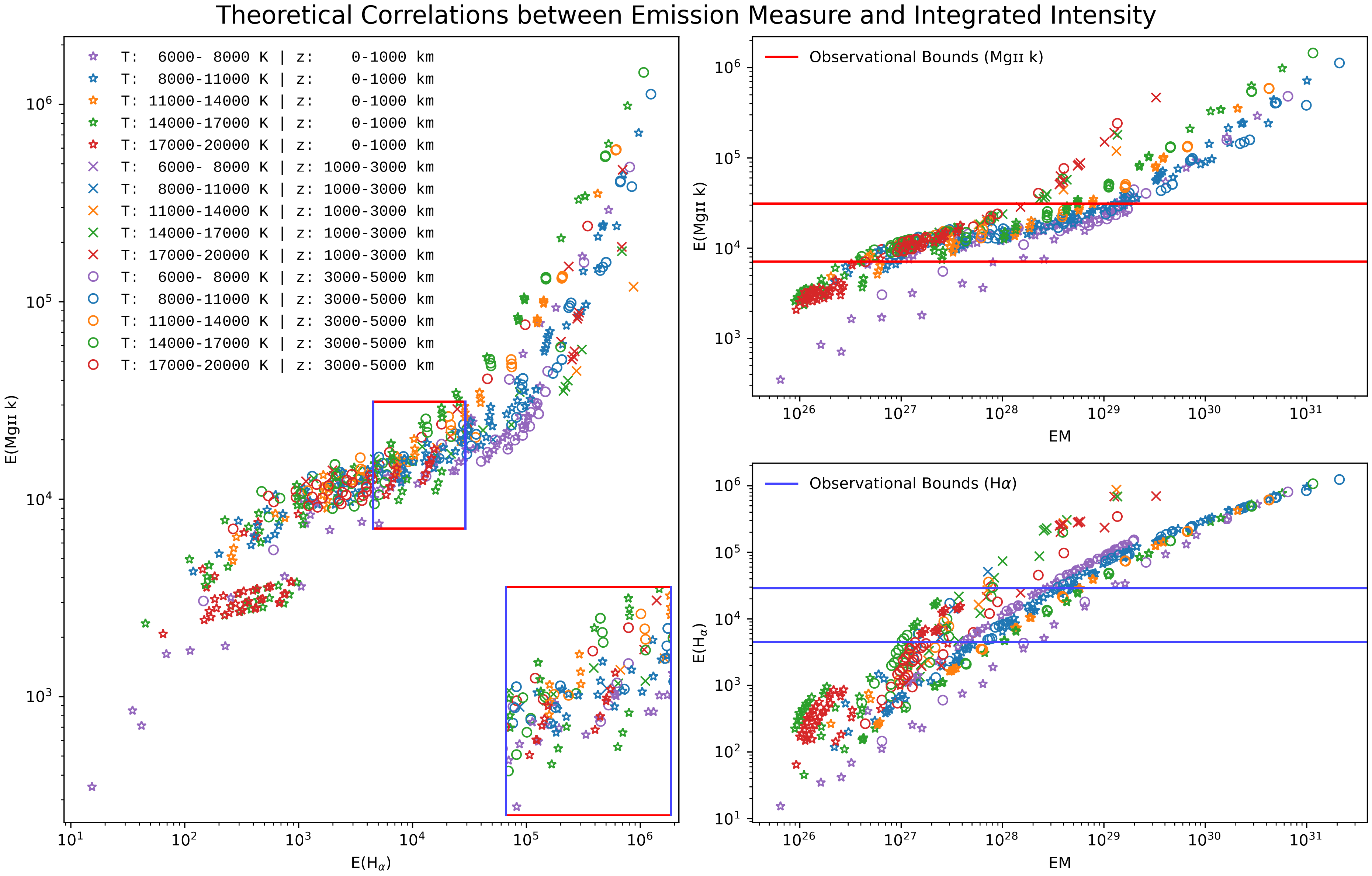}
\caption{Theoretical correlations
between emission measure and integrated intensity. (left) Relationship between the integrated intensity of \ion{Mg}{ii} k and the integrated intensity of H$\alpha$ derived from the theoretical models computed with a micro-turbulence equal to 5~km s$^{-1}$ and for different values of temperature and slab thickness. (top right) Integrated \ion{Mg}{ii} k intensity versus EM. (bottom right) Integrated H$\alpha$ intensity versus EM. The horizontal (vertical) lines limit the range of \ion{Mg}{ii} k (H$\alpha$) intensities. Integrated intensities are in erg s$^{-1}$ cm$^{-2}$ sr$^{-1}$ and EM are in cm$^{-5}$.
{ The small slab thickness models (stars) are new models. They fit quite well in the box of the observations.}
}\label{EM}
} 
\end{figure*}

\begin{figure*}
\centering
\includegraphics[width=0.8\textwidth]{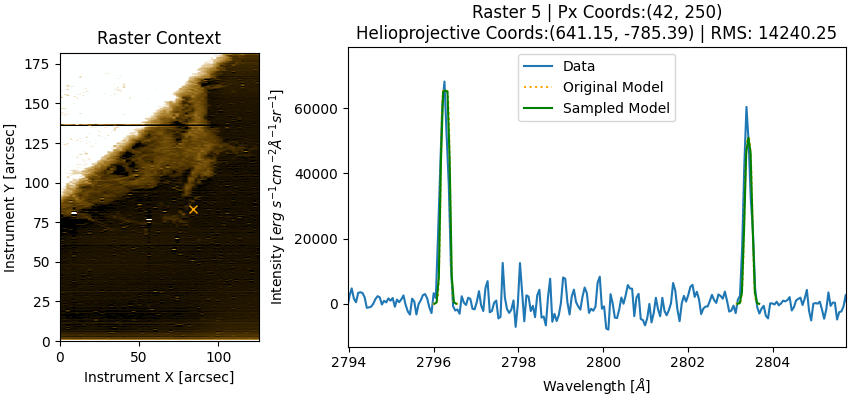}
\includegraphics[width=0.8\textwidth]{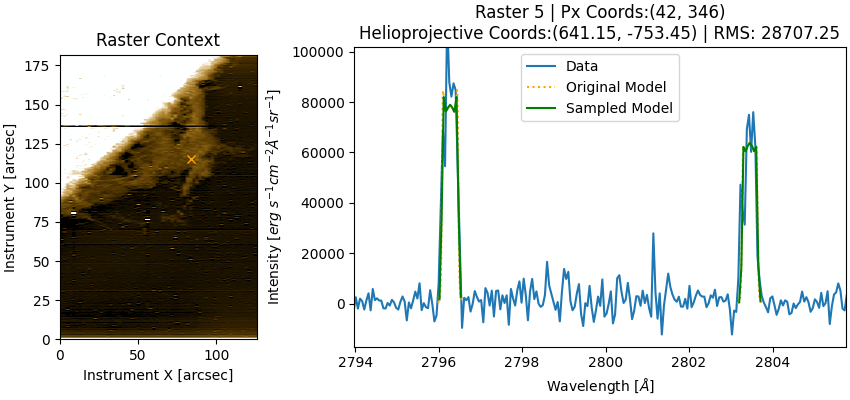}
\caption{Example of fitted theoretical profiles. The left shows an image of the prominence with an orange cross indicating the location of the profile \textit{Top:} A2, A profile in the horn well-fitted with one theoretical model. \textit{Bottom:} B11, Example of a profile in the prominence where the algorithm has attempted to fit a complex model to the complex structure found in this location. However, the RMS value is unsatisfactory.} \label{r5396}
\end{figure*}

\begin{figure*}[!ht]
\centering
\centering
\includegraphics[width=\textwidth]{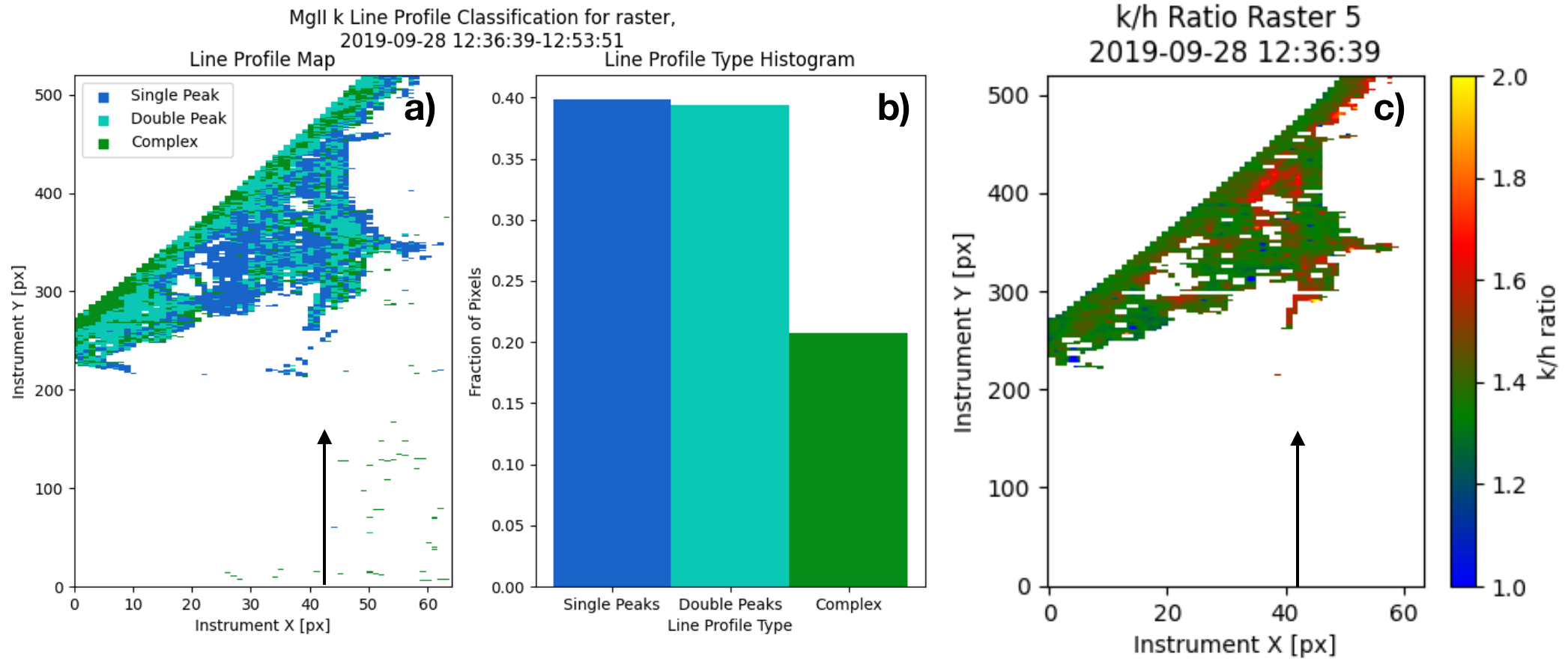}
\caption{Classification of the \ion{Mg}{ii} k line profiles in raster 5 (panels a and b), map of the intensity ratio of the \ion{Mg}{ii} k/h in raster 5 (panel c). { The coordinates of the maps are given in pixels to identify the selected points of Table \ref{tab:Obs} and Table \ref{tab:mg2_ha_sel} along x=42 px. The location x=42 px is marked with an arrow.}} \label{classification}
\end{figure*}

\section{Discussion and conclusion}\label{sec:dis_concl}

During a coordinated campaign between THEMIS and IRIS, we observed a very active  quiescent prominence on 28 September 2019 {with  small transient}  threads.  Two horns at the top of the prominence 
were well visible {for} one hour.  We focused our study on the horn and the edge of the prominence.
Using 23~940 NLTE 1D radiative transfer models we were able to fit the \ion{Mg}{ii} profiles in these thin structures, which mostly present one-peak Mg II profiles. }
We derived a few physical parameters such as electron density, mean temperature, {and} optical thickness.
The observed H$\alpha$ profiles from the coordinated campaign allowed us to compare the theoretical relationship between the integrated intensities of \ion{Mg}{ii} and H$\alpha$ and the emission measure (EM). 
Due to our selection of non-reversed \ion{Mg}{ii} profiles, our H$\alpha$ and \ion{Mg}{ii} profile analysis is restricted to the edge of the prominence. These observations behave in a similar manner to the relationship shown in Fig. \ref{EM}, but there is a large uncertainty in the determination of the parameters. In the future, we would like to analyse full prominence observations with 2D \ion{Mg}{ii} modelling.

{By performing pixel-by-pixel analyses on selected \ion{Mg}{ii}~h\&k and H$\alpha$ profiles in the edge of the prominence, and on selected \ion{Mg}{ii}~h\&k  profiles  in the horns},  we were able to derive the physical parameters of the plasma reliably.
 Overall, these two regions show similar ranges of values for the different parameters, but we note some trends -- namely  that more models with a PCTR are selected in the  horns, whereas  more isothermal and isobaric models fit the profiles in the edge region. We found that the electron density in the horns is slightly less than in the prominence column edge.
However, we  also find  a few pixels with relatively large electron density reaching nearly $2\times 10{^{10}}$~cm${^{-3}}$.
This has already been observed in tornado-like prominences and interpreted as plasma condensation \citep{Barczynski2021,Jenkins2022}. It was demonstrated that electron density could be high in fine structures at the prominence edges if narrow threads (of the order of 750~km) were considered \citep{Wiik1992}.

%

We { suggest}
that horns could be due to the LOS emission of the PCTR of prominence in dipped field lines, as proposed by \citet{Xia2014}. The density depletion in the cavity, visible in EUV and X-ray, is probably not due to a heating process of the plasma as suggested by \citet{Fan2019}. Instead, it may be attributed to the formation of plasma condensation or perhaps some drainage of material towards the chromosphere. The horn lifetime -- around 30m to 1h -- corresponds to the cooling time of coronal plasma and is therefore good support for this argument 
\citep{Schmieder1996}. The fine structures are very transient and could correspond to condensation in a weak twisted flux rope \citep{GuoJH2022}. { Before drawing a conclusion regarding the formation of the horns, we would require the full temporal evolution of the spectra.}

{Solar Orbiter remote-sensing instruments (EUI, SPICE) should be able to observe the horns of a prominence.}
In the future, the Chinese ASO- S satellite and onboard Lyman alpha telescope will be able to image these horns \citep{Zhao2022}. 2D radiative transfer code models should be able to provide a definitive answer on the formation of these horns in prominences \citep{Labrosse2016,Gunar2022}.\ Additionally, they may also be able to supply the amplitude and height of the heating influence, the topology of the fine structures, the density, and the temperature \citep{Jercic2022}.\

\begin{acknowledgements}
We thank the THEMIS team which assisted us in acquisition of the data during the coordinated THEMIS-IRIS campaign in September and October 2019.

IRIS is a NASA small explorer mission developed and operated by LMSAL with mission operations executed at NASA Ames Research Center and major contributions to downlink communications funded by ESA and the Norwegian Space Centre. AIA data courtesy of NASA/SDO and the AIA, EVE, and HMI science teams. Hinode is a Japanese mission developed and launched by ISAS/JAXA, collaborating with NAOJ as a domestic partner, NASA and UKSA as international partners. Scientific operation of the Hinode mission is conducted by the Hinode science team organized at ISAS/JAXA. This team mainly consists of scientists from institutes in the partner countries. Support for the post-launch operation is provided by JAXA and NAOJ (Japan), UKSA (U.K.), NASA, ESA, and NSC (Norway). The Global Oscillation Network Group (GONG) is a community-based program to conduct a detailed study of solar internal structure and dynamics using helioseismology. In order to exploit this new technique, GONG has developed a six-station network of extremely sensitive, and stable velocity imagers located around the Earth to obtain nearly continuous observations of the Sun's "five-minute" oscillations, or pulsations.

This study benefited from financial support from the Programme National Soleil Terre (PNST) of the CNRS/INSU, as well as from the Programme des Investissements d'Avenir (PIA) supervised by the ANR.
Support from STFC grants ST/S505390/1 (AWP) and ST/T000422/1 (NL) is gratefully acknowledged. Also thank you to Kathy Reeves for providing information on XRT data. This research used version 3.7.1 of Matplotlib \citep{matplotlib}, version 1.24.3 of NumPy \citep{harris_array_2020}, version 1.10.1 of SciPy \citep{scipy}, version 4.1.5 of the SunPy open source software package \citep{barnes_sunpy_2020}, and version 5.2.2 of Astropy (http://www.astropy.org) a community-developed core Python package for Astronomy \citep{the_astropy_collaboration_astropy_2013, the_astropy_collaboration_astropy_2018}

\end{acknowledgements}

\bibliographystyle{aa.bst} 
\bibliography{references.bib}

\begin{thebibliography}{64}
\expandafter\ifx\csname natexlab\endcsname\relax\def\natexlab#1{#1}\fi

\bibitem[{{Anzer} \& {Heinzel}(1999)}]{Anzer1999}
{Anzer}, U. \& {Heinzel}, P. 1999, \aap, 349, 974

\bibitem[{Aschwanden(2004)}]{aschwanden_physics_2004}
Aschwanden, M.~J. 2004, Physics of the solar corona: an introduction,
  Springer-{Praxis} books in geophysical sciences (Berlin ; New York: Springer)

\bibitem[{{Aulanier} \& {Schmieder}(2002)}]{Aulanier2002}
{Aulanier}, G. \& {Schmieder}, B. 2002, \aap, 386, 1106

\bibitem[{{Barczynski} {et~al.}(2021){Barczynski}, {Schmieder}, {Peat},
  {Labrosse}, {Mein}, \& {Mein}}]{Barczynski2021}
{Barczynski}, K., {Schmieder}, B., {Peat}, A.~W., {et~al.} 2021, \aap, 653, A94

\bibitem[{{Boggess} {et~al.}(1978){Boggess}, {Carr}, {Evans}, {Fischel},
  {Freeman}, {Fuechsel}, {Klinglesmith}, {Krueger}, {Longanecker}, \&
  {Moore}}]{iue1978}
{Boggess}, A., {Carr}, F.~A., {Evans}, D.~C., {et~al.} 1978, \nat, 275, 372

\bibitem[{{David}(1961)}]{David1961}
{David}, K.-H. 1961, Veroeffentlichungen der Universitaets-Sternwarte zu
  Goettingen, 7, 367

\bibitem[{{De Pontieu} {et~al.}(2014){De Pontieu}, {Title}, {Lemen}, {Kushner},
  {Akin}, {Allard}, {Berger}, {Boerner}, {Cheung}, {Chou}, {Drake}, {Duncan},
  {Freeland}, {Heyman}, {Hoffman}, {Hurlburt}, {Lindgren}, {Mathur}, {Rehse},
  {Sabolish}, {Seguin}, {Schrijver}, {Tarbell}, {W{\"u}lser}, {Wolfson},
  {Yanari}, {Mudge}, {Nguyen-Phuc}, {Timmons}, {van Bezooijen}, {Weingrod},
  {Brookner}, {Butcher}, {Dougherty}, {Eder}, {Knagenhjelm}, {Larsen},
  {Mansir}, {Phan}, {Boyle}, {Cheimets}, {DeLuca}, {Golub}, {Gates}, {Hertz},
  {McKillop}, {Park}, {Perry}, {Podgorski}, {Reeves}, {Saar}, {Testa}, {Tian},
  {Weber}, {Dunn}, {Eccles}, {Jaeggli}, {Kankelborg}, {Mashburn}, {Pust},
  {Springer}, {Carvalho}, {Kleint}, {Marmie}, {Mazmanian}, {Pereira}, {Sawyer},
  {Strong}, {Worden}, {Carlsson}, {Hansteen}, {Leenaarts}, {Wiesmann},
  {Aloise}, {Chu}, {Bush}, {Scherrer}, {Brekke}, {Martinez-Sykora}, {Lites},
  {McIntosh}, {Uitenbroek}, {Okamoto}, {Gummin}, {Auker}, {Jerram}, {Pool}, \&
  {Waltham}}]{DePontieu2014}
{De Pontieu}, B., {Title}, A.~M., {Lemen}, J.~R., {et~al.} 2014, \solphys, 289,
  2733

\bibitem[{{Fan} \& {Liu}(2019)}]{Fan2019}
{Fan}, Y. \& {Liu}, T. 2019, Frontiers in Astronomy and Space Sciences, 6, 27

\bibitem[{{Freeland} \& {Handy}(1998)}]{freeland1998}
{Freeland}, S.~L. \& {Handy}, B.~N. 1998, \solphys, 182, 497

\bibitem[{{Gibson}(2015)}]{Gibson2015}
{Gibson}, S. 2015, in Astrophysics and Space Science Library, Vol. 415, Solar
  Prominences, ed. J.-C. {Vial} \& O.~{Engvold}, 323

\bibitem[{{Gibson}(2018)}]{Gibson2018}
{Gibson}, S.~E. 2018, Living Reviews in Solar Physics, 15, 7

\bibitem[{Golub {et~al.}(2007)Golub, DeLuca, Austin, Bookbinder, Caldwell,
  Cheimets, Cirtain, Cosmo, Reid, Sette, Weber, Sakao, Kano, Shibasaki, Hara,
  Tsuneta, Kumagai, Tamura, Shimojo, McCracken, Carpenter, Haight, Siler,
  Wright, Tucker, Rutledge, Barbera, Peres, \& Varisco}]{golub_x-ray_2007}
Golub, L., DeLuca, E., Austin, G., {et~al.} 2007, Solar Physics, 243, 63

\bibitem[{{Gouttebroze} {et~al.}(1993){Gouttebroze}, {Heinzel}, \&
  {Vial}}]{Gouutebroze1993}
{Gouttebroze}, P., {Heinzel}, P., \& {Vial}, J.~C. 1993, \aaps, 99, 513

\bibitem[{{Gouttebroze} {et~al.}(1997){Gouttebroze}, {Vial}, \&
  {Heinzel}}]{gouttebroze_1997}
{Gouttebroze}, P., {Vial}, J.~C., \& {Heinzel}, P. 1997, \solphys, 172, 125

\bibitem[{{Gun{\'a}r} {et~al.}(2018){Gun{\'a}r}, {Dud{\'\i}k}, {Aulanier},
  {Schmieder}, \& {Heinzel}}]{Gunar2018}
{Gun{\'a}r}, S., {Dud{\'\i}k}, J., {Aulanier}, G., {Schmieder}, B., \&
  {Heinzel}, P. 2018, \apj, 867, 115

\bibitem[{{Gun{\'a}r} {et~al.}(2022){Gun{\'a}r}, {Heinzel}, {Koza}, \&
  {Schwartz}}]{Gunar2022}
{Gun{\'a}r}, S., {Heinzel}, P., {Koza}, J., \& {Schwartz}, P. 2022, \apj, 934,
  133

\bibitem[{{Guo} {et~al.}(2022){Guo}, {Ni}, {Zhou}, {Guo}, {Schmieder}, \&
  {Chen}}]{GuoJH2022}
{Guo}, J.~H., {Ni}, Y.~W., {Zhou}, Y.~H., {et~al.} 2022, \aap, 667, A89

\bibitem[{Harris {et~al.}(2020)Harris, Millman, van~der Walt, Gommers,
  Virtanen, Cournapeau, Wieser, Taylor, Berg, Smith, Kern, Picus, Hoyer, van
  Kerkwijk, Brett, Haldane, del R{\'i}o, Wiebe, Peterson, G{\'e}rard-Marchant,
  Sheppard, Reddy, Weckesser, Abbasi, Gohlke, \& Oliphant}]{harris_array_2020}
Harris, C.~R., Millman, K.~J., van~der Walt, S.~J., {et~al.} 2020, Nature, 585,
  357

\bibitem[{{Harvey} {et~al.}(1996){Harvey}, {Hill}, {Hubbard}, {Kennedy},
  {Leibacher}, {Pintar}, {Gilman}, {Noyes}, {Title}, {Toomre}, {Ulrich},
  {Bhatnagar}, {Kennewell}, {Marquette}, {Patron}, {Saa}, \& {Yasukawa}}]{gong}
{Harvey}, J.~W., {Hill}, F., {Hubbard}, R.~P., {et~al.} 1996, Science, 272,
  1284

\bibitem[{{Heinzel} {et~al.}(1994){Heinzel}, {Gouttebroze}, \&
  {Vial}}]{heinzel1994}
{Heinzel}, P., {Gouttebroze}, P., \& {Vial}, J.~C. 1994, \aap, 292, 656

\bibitem[{{Heinzel} {et~al.}(2015){Heinzel}, {Schmieder}, {Mein}, \&
  {Gun{\'a}r}}]{Heinzel2015}
{Heinzel}, P., {Schmieder}, B., {Mein}, N., \& {Gun{\'a}r}, S. 2015, \apjl,
  800, L13

\bibitem[{{Heinzel} {et~al.}(2001){Heinzel}, {Schmieder}, \&
  {Tziotziou}}]{heinzel2001b}
{Heinzel}, P., {Schmieder}, B., \& {Tziotziou}, K. 2001, \apjl, 561, L223

\bibitem[{{Heinzel} {et~al.}(2014){Heinzel}, {Vial}, \& {Anzer}}]{Heinzel2014}
{Heinzel}, P., {Vial}, J.-C., \& {Anzer}, U. 2014, \aap, 564, A132

\bibitem[{Hunter(2007)}]{matplotlib}
Hunter, J.~D. 2007, Comp. Sci, Eng., 9, 90

\bibitem[{{Jej{\v{c}}i{\v{c}}} {et~al.}(2022){Jej{\v{c}}i{\v{c}}}, {Heinzel},
  {Schmieder}, {Gun{\'a}r}, {Mein}, {Mein}, \& {Ruan}}]{Jejcic2022}
{Jej{\v{c}}i{\v{c}}}, S., {Heinzel}, P., {Schmieder}, B., {et~al.} 2022, \apj,
  932, 3

\bibitem[{{Jej{\v{c}}i{\v{c}}} {et~al.}(2018){Jej{\v{c}}i{\v{c}}}, {Schwartz},
  {Heinzel}, {Zapi{\'o}r}, \& {Gun{\'a}r}}]{Jejcic2018}
{Jej{\v{c}}i{\v{c}}}, S., {Schwartz}, P., {Heinzel}, P., {Zapi{\'o}r}, M., \&
  {Gun{\'a}r}, S. 2018, \aap, 618, A88

\bibitem[{{Jenkins} \& {Keppens}(2022)}]{Jenkins2022}
{Jenkins}, J.~M. \& {Keppens}, R. 2022, Nature Astronomy, 6, 942

\bibitem[{{Jer{\v{c}}i{\'c}} {et~al.}(2022){Jer{\v{c}}i{\'c}}, {Keppens}, \&
  {Zhou}}]{Jercic2022}
{Jer{\v{c}}i{\'c}}, V., {Keppens}, R., \& {Zhou}, Y. 2022, \aap, 658, A58

\bibitem[{{Kerr} {et~al.}(2015){Kerr}, {Sim{\~o}es}, {Qiu}, \&
  {Fletcher}}]{Kerr2015}
{Kerr}, G.~S., {Sim{\~o}es}, P.~J.~A., {Qiu}, J., \& {Fletcher}, L. 2015, \aap,
  582, A50

\bibitem[{Kosugi {et~al.}(2007)Kosugi, Matsuzaki, Sakao, Shimizu, Sone,
  Tachikawa, Hashimoto, Minesugi, Ohnishi, Yamada, Tsuneta, Hara, Ichimoto,
  Suematsu, Shimojo, Watanabe, Shimada, Davis, Hill, Owens, Title, Culhane,
  Harra, Doschek, \& Golub}]{kosugi_hinode_2007}
Kosugi, T., Matsuzaki, K., Sakao, T., {et~al.} 2007, Solar Physics, 243, 3

\bibitem[{{Labrosse} \& {Gouttebroze}(2004)}]{labrosse2004}
{Labrosse}, N. \& {Gouttebroze}, P. 2004, \apj, 617, 614

\bibitem[{Labrosse {et~al.}(2010)Labrosse, Heinzel, Vial, Kucera, Parenti,
  Gun\'{a}r, Schmieder, \& Kilper}]{Labrosse2010}
Labrosse, N., Heinzel, P., Vial, J.-C., {et~al.} 2010, Space Sci. Rev., 151,
  243

\bibitem[{{Labrosse} \& {Rodger}(2016)}]{Labrosse2016}
{Labrosse}, N. \& {Rodger}, A.~S. 2016, \aap, 587, A113

\bibitem[{{Lemen} {et~al.}(2012){Lemen}, {Title}, {Akin}, {Boerner}, {Chou},
  {Drake}, {Duncan}, {Edwards}, {Friedlaender}, {Heyman}, {Hurlburt}, {Katz},
  {Kushner}, {Levay}, {Lindgren}, {Mathur}, {McFeaters}, {Mitchell}, {Rehse},
  {Schrijver}, {Springer}, {Stern}, {Tarbell}, {Wuelser}, {Wolfson}, {Yanari},
  {Bookbinder}, {Cheimets}, {Caldwell}, {Deluca}, {Gates}, {Golub}, {Park},
  {Podgorski}, {Bush}, {Scherrer}, {Gummin}, {Smith}, {Auker}, {Jerram},
  {Pool}, {Soufli}, {Windt}, {Beardsley}, {Clapp}, {Lang}, \&
  {Waltham}}]{Lemen2012}
{Lemen}, J.~R., {Title}, A.~M., {Akin}, D.~J., {et~al.} 2012, \solphys, 275, 17

\bibitem[{{Levens} \& {Labrosse}(2019)}]{Levens2019}
{Levens}, P.~J. \& {Labrosse}, N. 2019, \aap, 625, A30

\bibitem[{{Luna} {et~al.}(2012){Luna}, {Karpen}, \& {DeVore}}]{Luna2012}
{Luna}, M., {Karpen}, J.~T., \& {DeVore}, C.~R. 2012, \apj, 746, 30

\bibitem[{{Mackay} {et~al.}(2010){Mackay}, {Karpen}, {Ballester}, {Schmieder},
  \& {Aulanier}}]{Mackay2010}
{Mackay}, D.~H., {Karpen}, J.~T., {Ballester}, J.~L., {Schmieder}, B., \&
  {Aulanier}, G. 2010, \ssr, 151, 333

\bibitem[{Mein \& Rayrole(1985)}]{mein_themis_1985}
Mein, P. \& Rayrole, J. 1985, Vistas in Astronomy, 28, 567

\bibitem[{{Moore} {et~al.}(1966){Moore}, {Minnaert}, \& {Houtgast}}]{Moore1966}
{Moore}, C.~E., {Minnaert}, M.~G.~J., \& {Houtgast}, J. 1966, {The solar
  spectrum 2935 A to 8770 A}

\bibitem[{{Peat}(2023)}]{Peat2023}
{Peat}, A.~W. 2023, PhD thesis, University of Glasgow, UK

\bibitem[{{Peat} {et~al.}(in prep.){Peat}, {Labrosse}, {Barczynski}, \&
  {Schmieder}}]{peatprep}
{Peat}, A.~W., {Labrosse}, N., {Barczynski}, K., \& {Schmieder}, B. in prep.

\bibitem[{{Peat} {et~al.}(2021){Peat}, {Labrosse}, {Schmieder}, \&
  {Barczynski}}]{Peat2021}
{Peat}, A.~W., {Labrosse}, N., {Schmieder}, B., \& {Barczynski}, K. 2021, \aap,
  653, A5

\bibitem[{{Pesnell} {et~al.}(2012){Pesnell}, {Thompson}, \&
  {Chamberlin}}]{Pesnell2012}
{Pesnell}, W.~D., {Thompson}, B.~J., \& {Chamberlin}, P.~C. 2012, \solphys,
  275, 3

\bibitem[{{Ruan} {et~al.}(2019){Ruan}, {Jej{\v{c}}i{\v{c}}}, {Schmieder},
  {Mein}, {Mein}, {Heinzel}, {Gun{\'a}r}, \& {Chen}}]{Ruan2019}
{Ruan}, G., {Jej{\v{c}}i{\v{c}}}, S., {Schmieder}, B., {et~al.} 2019, \apj,
  886, 134

\bibitem[{{Ruan} {et~al.}(2018){Ruan}, {Schmieder}, {Mein}, {Mein}, {Labrosse},
  {Gun{\'a}r}, \& {Chen}}]{Ruan2018}
{Ruan}, G., {Schmieder}, B., {Mein}, P., {et~al.} 2018, \apj, 865, 123

\bibitem[{{Schmieder} {et~al.}(2010){Schmieder}, {Chandra}, {Berlicki}, \&
  {Mein}}]{Schmieder2010}
{Schmieder}, B., {Chandra}, R., {Berlicki}, A., \& {Mein}, P. 2010, \aap, 514,
  A68

\bibitem[{{Schmieder} {et~al.}(1996){Schmieder}, {Heinzel}, {Wiik}, {Lemen}, \&
  {Hiei}}]{Schmieder1996}
{Schmieder}, B., {Heinzel}, P., {Wiik}, J.~E., {Lemen}, J., \& {Hiei}, E. 1996,
  Advances in Space Research, 17, 111

\bibitem[{{Schmieder} {et~al.}(2017){Schmieder}, {Mein}, {Mein}, {Levens},
  {Labrosse}, \& {Ofman}}]{Schmieder2017}
{Schmieder}, B., {Mein}, P., {Mein}, N., {et~al.} 2017, \aap, 597, A109

\bibitem[{{Schmieder} {et~al.}(2014){Schmieder}, {Tian}, {Kucera}, {L{\'o}pez
  Ariste}, {Mein}, {Mein}, {Dalmasse}, \& {Golub}}]{Schmieder2004b}
{Schmieder}, B., {Tian}, H., {Kucera}, T., {et~al.} 2014, \aap, 569, A85

\bibitem[{{Schmit} \& {Gibson}(2013)}]{Schmit2013a}
{Schmit}, D.~J. \& {Gibson}, S. 2013, \apj, 770, 35

\bibitem[{{Schmit} {et~al.}(2013){Schmit}, {Gibson}, {Luna}, {Karpen}, \&
  {Innes}}]{Schmit2013b}
{Schmit}, D.~J., {Gibson}, S., {Luna}, M., {Karpen}, J., \& {Innes}, D. 2013,
  \apj, 779, 156

\bibitem[{{SunPy Community} {et~al.}(2020){SunPy Community}, {Barnes}, {Bobra},
  {Christe}, {Freij}, {Hayes}, {Ireland}, {Mumford}, {Perez-Suarez}, {Ryan},
  {Shih}, {Chanda}, {Glogowski}, {Hewett}, {Hughitt}, {Hill}, {Hiware},
  {Inglis}, {Kirk}, {Konge}, {Mason}, {Maloney}, {Murray}, {Panda}, {Park},
  {Pereira}, {Reardon}, {Savage}, {Sip{\H{o}}cz}, {Stansby}, {Jain}, {Taylor},
  {Yadav}, {Rajul}, \& {Dang}}]{barnes_sunpy_2020}
{SunPy Community}, {Barnes}, W.~T., {Bobra}, M.~G., {et~al.} 2020, \apj, 890,
  68

\bibitem[{{Tandberg-Hanssen}(1995)}]{th1995}
{Tandberg-Hanssen}, E. 1995, {The nature of solar prominences}, Vol. 199

\bibitem[{{The Astropy Collaboration} {et~al.}(2018){The Astropy
  Collaboration}, Price-Whelan, Sip{\H o}cz, G{\"u}nther, Lim, Crawford,
  Conseil, Shupe, Craig, Dencheva, Ginsburg, VanderPlas, Bradley,
  P{\'e}rez-Su{\'a}rez, de~Val-Borro, {(Primary Paper Contributors)}, Aldcroft,
  Cruz, Robitaille, Tollerud, {(Astropy Coordination Committee)}, Ardelean,
  Babej, Bach, Bachetti, Bakanov, Bamford, Barentsen, Barmby, Baumbach, Berry,
  Biscani, Boquien, Bostroem, Bouma, Brammer, Bray, Breytenbach, Buddelmeijer,
  Burke, Calderone, Rodr{\'i}guez, Cara, Cardoso, Cheedella, Copin, Corrales,
  Crichton, D{\textquoteright}Avella, Deil, Depagne, Dietrich, Donath,
  Droettboom, Earl, Erben, Fabbro, Ferreira, Finethy, Fox, Garrison, Gibbons,
  Goldstein, Gommers, Greco, Greenfield, Groener, Grollier, Hagen, Hirst,
  Homeier, Horton, Hosseinzadeh, Hu, Hunkeler, Ivezi{\'c}, Jain, Jenness,
  Kanarek, Kendrew, Kern, Kerzendorf, Khvalko, King, Kirkby, Kulkarni, Kumar,
  Lee, Lenz, Littlefair, Ma, Macleod, Mastropietro, McCully, Montagnac, Morris,
  Mueller, Mumford, Muna, Murphy, Nelson, Nguyen, Ninan, N{\"o}the, Ogaz, Oh,
  Parejko, Parley, Pascual, Patil, Patil, Plunkett, Prochaska, Rastogi, Janga,
  Sabater, Sakurikar, Seifert, Sherbert, Sherwood-Taylor, Shih, Sick, Silbiger,
  Singanamalla, Singer, Sladen, Sooley, Sornarajah, Streicher, Teuben, Thomas,
  Tremblay, Turner, Terr{\'o}n, Kerkwijk, de~la Vega, Watkins, Weaver,
  Whitmore, Woillez, Zabalza, \& {(Astropy
  Contributors)}}]{the_astropy_collaboration_astropy_2018}
{The Astropy Collaboration}, Price-Whelan, A.~M., Sip{\H o}cz, B.~M., {et~al.}
  2018, The Astronomical Journal, 156, 123

\bibitem[{{The Astropy Collaboration} {et~al.}(2013){The Astropy
  Collaboration}, Robitaille, Tollerud, Greenfield, Droettboom, Bray, Aldcroft,
  Davis, Ginsburg, Price-Whelan, Kerzendorf, Conley, Crighton, Barbary, Muna,
  Ferguson, Grollier, Parikh, Nair, G{\"u}nther, Deil, Woillez, Conseil,
  Kramer, Turner, Singer, Fox, Weaver, Zabalza, Edwards, Azalee~Bostroem,
  Burke, Casey, Crawford, Dencheva, Ely, Jenness, Labrie, Lim, Pierfederici,
  Pontzen, Ptak, Refsdal, Servillat, \&
  Streicher}]{the_astropy_collaboration_astropy_2013}
{The Astropy Collaboration}, Robitaille, T.~P., Tollerud, E.~J., {et~al.} 2013,
  Astronomy \& Astrophysics, 558, A33

\bibitem[{Tian {et~al.}(2014)Tian, De~Pontieu, DeLuca, Weulser, \&
  Testa}]{tian_itn_2014}
Tian, H., De~Pontieu, B., DeLuca, E., Weulser, J.-P., \& Testa, P. 2014, {ITN}
  24: Stellar Calibration

\bibitem[{{Vial}(1998)}]{vial1998}
{Vial}, J.~C. 1998, in Astronomical Society of the Pacific Conference Series,
  Vol. 150, IAU Colloq. 167: New Perspectives on Solar Prominences, ed. D.~F.
  {Webb}, B.~{Schmieder}, \& D.~M. {Rust}, 175

\bibitem[{{Vial} {et~al.}(2019){Vial}, {Zhang}, \& {Buchlin}}]{Vial2019}
{Vial}, J.~C., {Zhang}, P., \& {Buchlin}, {\'E}. 2019, \aap, 624, A56

\bibitem[{Virtanen {et~al.}(2020)Virtanen, Gommers, Oliphant, Haberland, Reddy,
  Cournapeau, Burovski, Peterson, Weckesser, Bright, {van der Walt}, Brett,
  Wilson, Millman, Mayorov, Nelson, Jones, Kern, Larson, Carey, Polat, Feng,
  Moore, {VanderPlas}, Laxalde, Perktold, Cimrman, Henriksen, Quintero, Harris,
  Archibald, Ribeiro, Pedregosa, {van Mulbregt}, \& {SciPy 1.0
  Contributors}}]{scipy}
Virtanen, P., Gommers, R., Oliphant, T.~E., {et~al.} 2020, Nat. Methods, 17,
  261

\bibitem[{{Wang} {et~al.}(2016){Wang}, {Chen}, {Fu}, {Li}, {Li}, \&
  {Liu}}]{Wang2016}
{Wang}, B., {Chen}, Y., {Fu}, J., {et~al.} 2016, \apjl, 827, L33

\bibitem[{{Wiik} {et~al.}(1992){Wiik}, {Heinzel}, \& {Schmieder}}]{Wiik1992}
{Wiik}, J.~E., {Heinzel}, P., \& {Schmieder}, B. 1992, \aap, 260, 419

\bibitem[{{Xia} {et~al.}(2014){Xia}, {Keppens}, {Antolin}, \&
  {Porth}}]{Xia2014}
{Xia}, C., {Keppens}, R., {Antolin}, P., \& {Porth}, O. 2014, \apjl, 792, L38

\bibitem[{{Zhang} {et~al.}(2019){Zhang}, {Buchlin}, \& {Vial}}]{Zhang2019}
{Zhang}, P., {Buchlin}, {\'E}., \& {Vial}, J.~C. 2019, \aap, 624, A72

\bibitem[{{Zhao} {et~al.}(2022){Zhao}, {Zhang}, {Gibson}, {Fan}, {Feng}, {Yu},
  {Li}, \& {Gan}}]{Zhao2022}
{Zhao}, J., {Zhang}, P., {Gibson}, S.~E., {et~al.} 2022, \aap, 665, A39

\end{thebibliography}

\begin{appendix}
\section{Prominence evolution}\label{ap:ap1}
Figures~\ref{AIA_304}, \ref{AIA_171}, and \ref{AIA_193} are snapshots of the AIA 304~\AA{}, AIA 171\AA, and AIA 193~\AA{}\ movies. They show the evolution of the prominence with the two horns at the top.
The contour of the prominence with horns obtained from the AIA 304~\AA{}\ map at 12:29 UT is overlaid in the AIA 171~\AA\ and, AIA 193~\AA{}\ maps (panels c) to clearly identify the horns, which are also well observed in the panels (b).

\begin{figure*}[!htb]
\centering
\includegraphics[width=18cm]{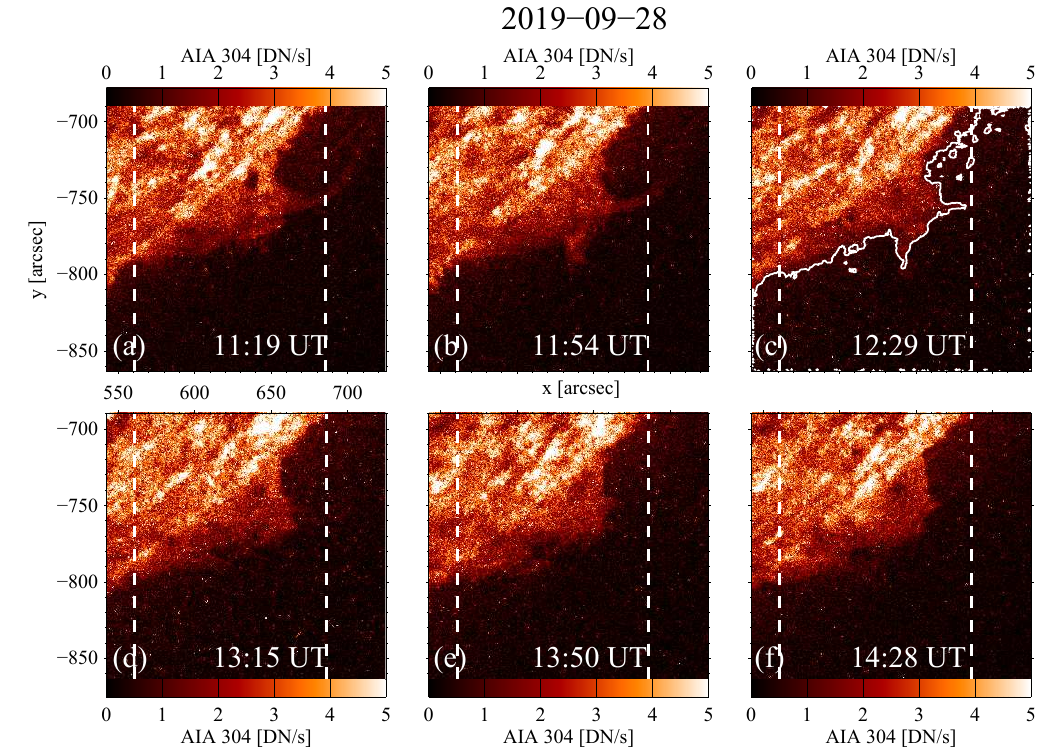}
\caption{AIA 304 images; snapshots of the movie (AIA304.mp4). The white line marks the raster position.The contour of the AIA 304~\AA{}\ is overlaid in panel (c) to show the horns at the top of the prominence. \label{AIA_304}}
\end{figure*}

\begin{figure*}[!htb]
\centering
\includegraphics[width=18cm]{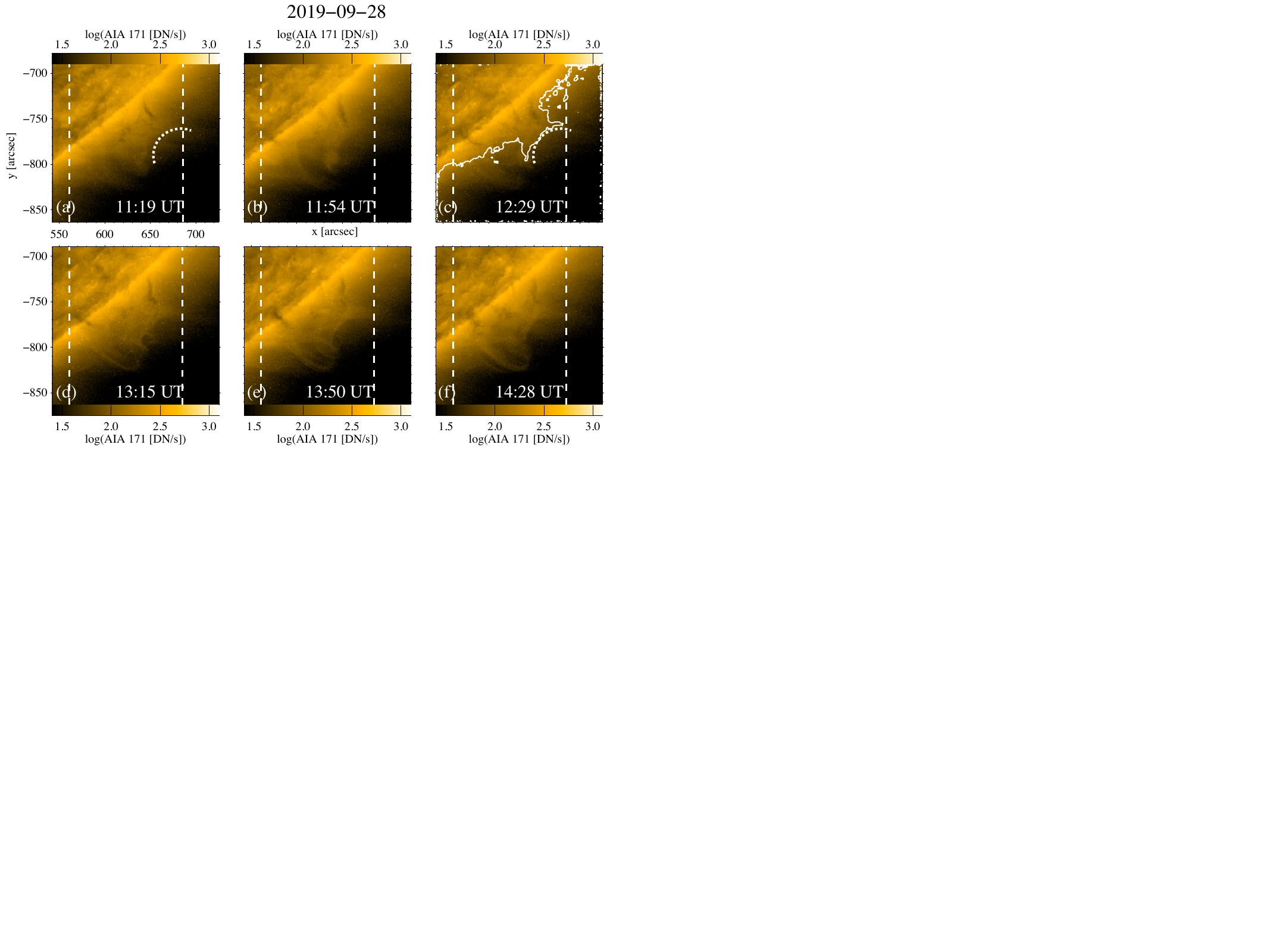}
\caption{AIA 171 images; snapshots of the movie (AIA171.mp4). The white line marks the raster position. The contour of the AIA 304~\AA{}\ is overlaid in panel (c) to show the horns at the top of the prominence. The cavity is marked with a white dotted line. \label{AIA_171}}
\end{figure*}

\begin{figure*}[!htb]
\centering
\includegraphics[width=18cm]{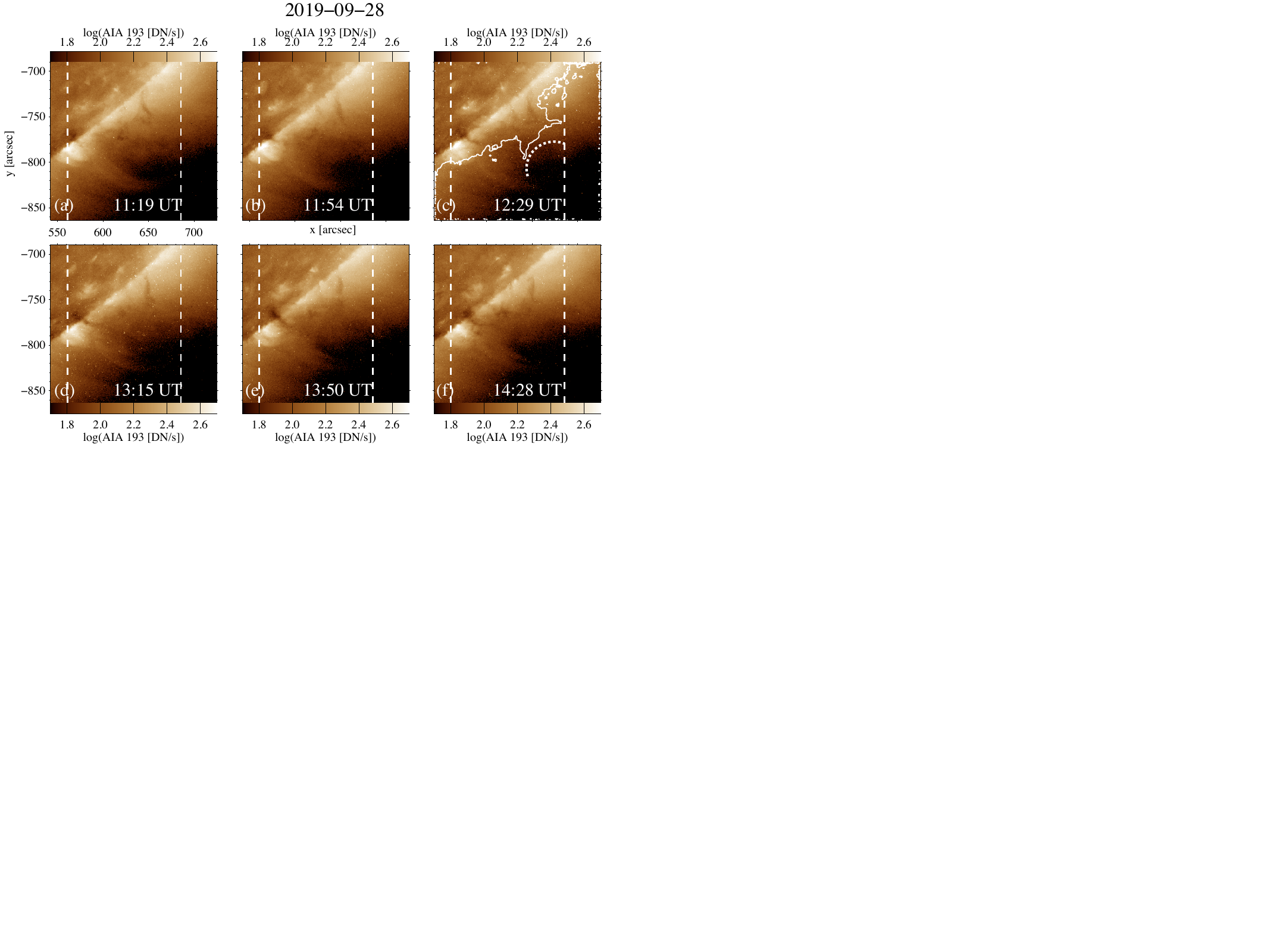}
\caption{AIA 193 images; snapshots of the movie (AIA193.mp4). The white line marks the raster position. The contour of the AIA 304~\AA{}\ is overlaid in panel (c) to show the horns at the top of the prominence. The cavity is marked with a white dotted line.} \label{AIA_193}
\end{figure*}

\section{Calibration}\label{ap:ap2}
The Multi Thermal Raie mode (MTR) of THEMIS observes in a large wavelength domain (6\AA), which allows us to almost reach the continuum in the environment of the H$\alpha$ line (Figure~\ref{fig:calibration}). This permits us to calibrate the wavelength domain, to correct the scattering light, and to have relatively precise measurements of the FWHM. This is not the case with other instruments such as the SST, which observes with filters ($\pm$1.5~\AA{}) or with the MSDP ($\pm$0.7~\AA{}).

To calibrate the H$\alpha$ line in wavelength, we used two telluric lines of H$_{2}$O located in the spectrum on the right (line A, 6560.555\AA) and left (line B, 6564.206\AA) sides of H$\alpha$.
We chose the ROI located at the solar disc.
For each telluric line, we defined a band that covers the whole line, and we fitted a Gaussian function with a slope and an offset \citep{Moore1966}.


The H$\alpha$ intensity was calibrated by fitting the profiles observed at different $\mu$ values on the disc with the David profile, taking into account the limb darkening and the position on the disc. The value of the continuum intensity close to the H$\alpha$ line at the disc centre is 
4.077 $\times 10^{-5} $ erg cm$^{-2}$ s$^{-1}$ sr$^{-1}$ Hz$^{-1}$.

\begin{figure*}[!htb]
\centering
\includegraphics[width=\textwidth]{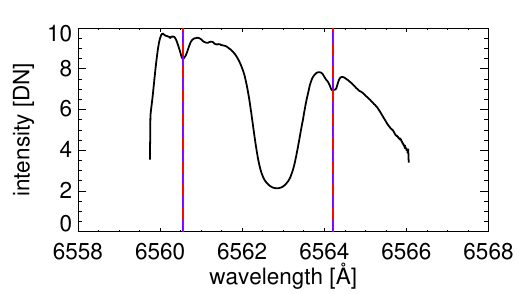}
\caption{Mean H$\alpha$ line profile obtained on disc at $\mu= 0.99$ along a circular slit used to fit with the David profile for the intensity calibration. The vertical red lines indicate the position of the telluric lines for the wavelength calibration. \label{fig:calibration}}
\end{figure*}

\section{Co-temporal and co-spatial H$\alpha$ and \ion{Mg}{ii} line profiles}\label{ap:ap3}
Figures~\ref{fig:IRIS_A}, \ref{fig:IRIS_B}, and \ref{Ha_B} present the series of profiles corresponding to pixels enumerated in Table~\ref{tab:mg2_ha_sel}.
H$\alpha$ profiles are co-temporal and co-spatial of IRIS profiles in the B series. 
There are no H$\alpha$ profiles corresponding to the A IRIS series because there are no counts.

\begin{figure*}[!htb]
\centering
\includegraphics[width=\textwidth]{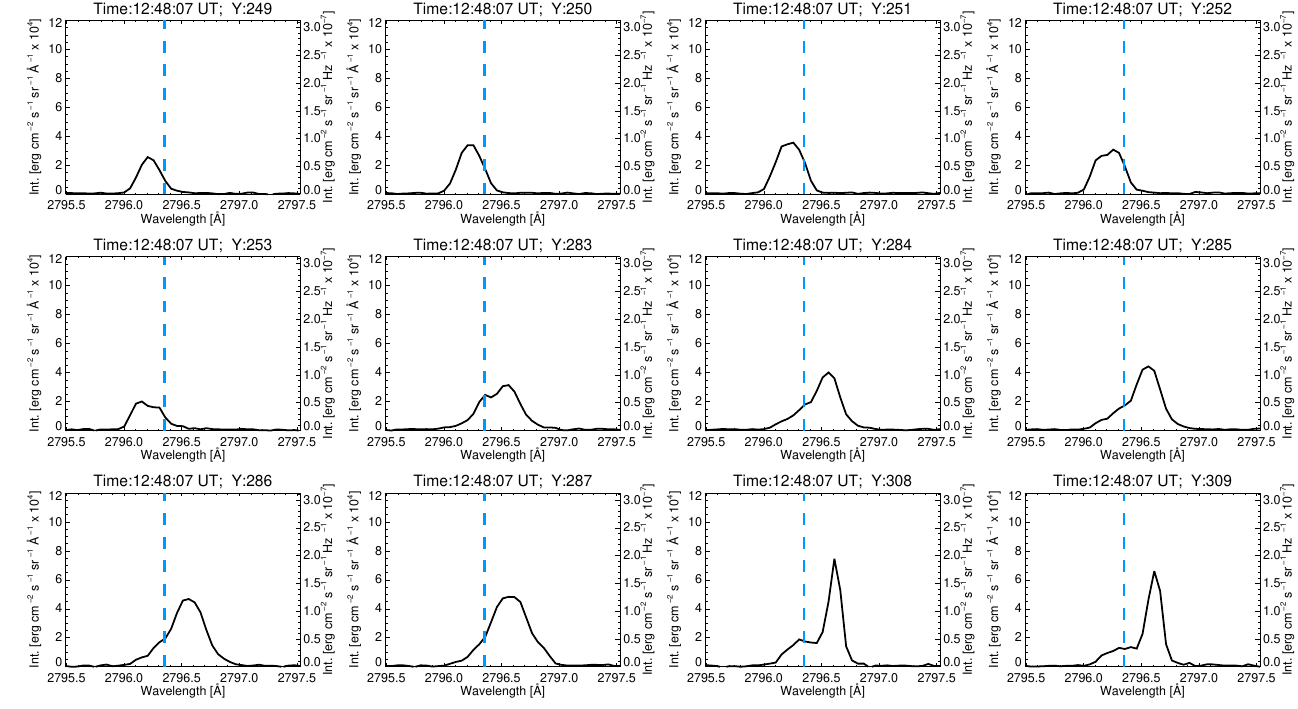}
\caption{IRIS: \ion{Mg}{ii} k along slit 42 in raster 5 in the area of the horns (pixels A in Table~\ref{tab:mg2_ha_sel}). \label{fig:IRIS_A}}
\end{figure*}

\begin{figure*}[!htb]
\centering
\includegraphics[width=\textwidth]{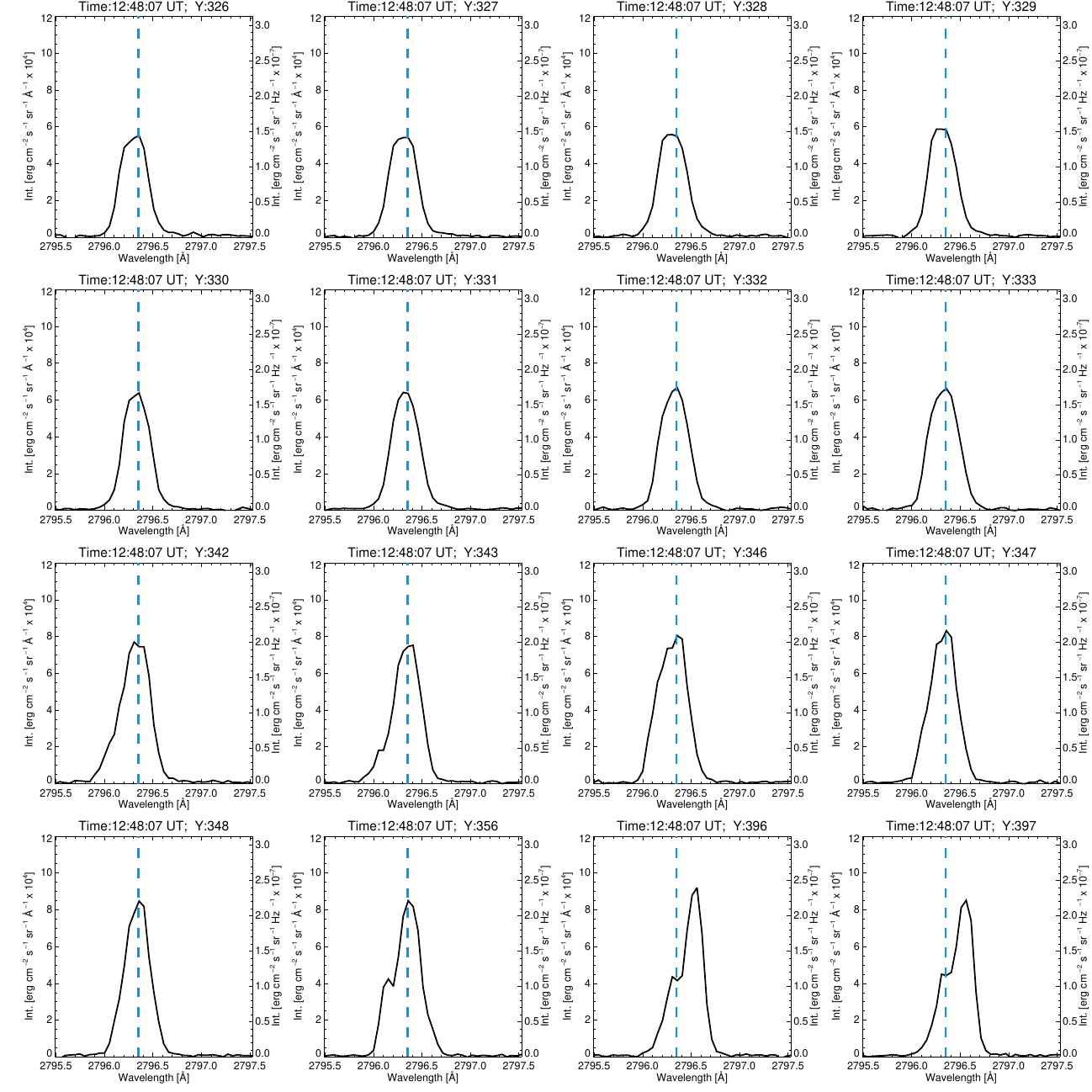}
\caption{IRIS,:\ion{Mg}{ii} k along slit 42 in raster 5 in the area at the edge of the prominence column (pixels B Table~\ref{tab:mg2_ha_sel}). \label{fig:IRIS_B}}
\end{figure*}

\begin{figure*}[!htb]
\centering
\includegraphics[width=\textwidth]{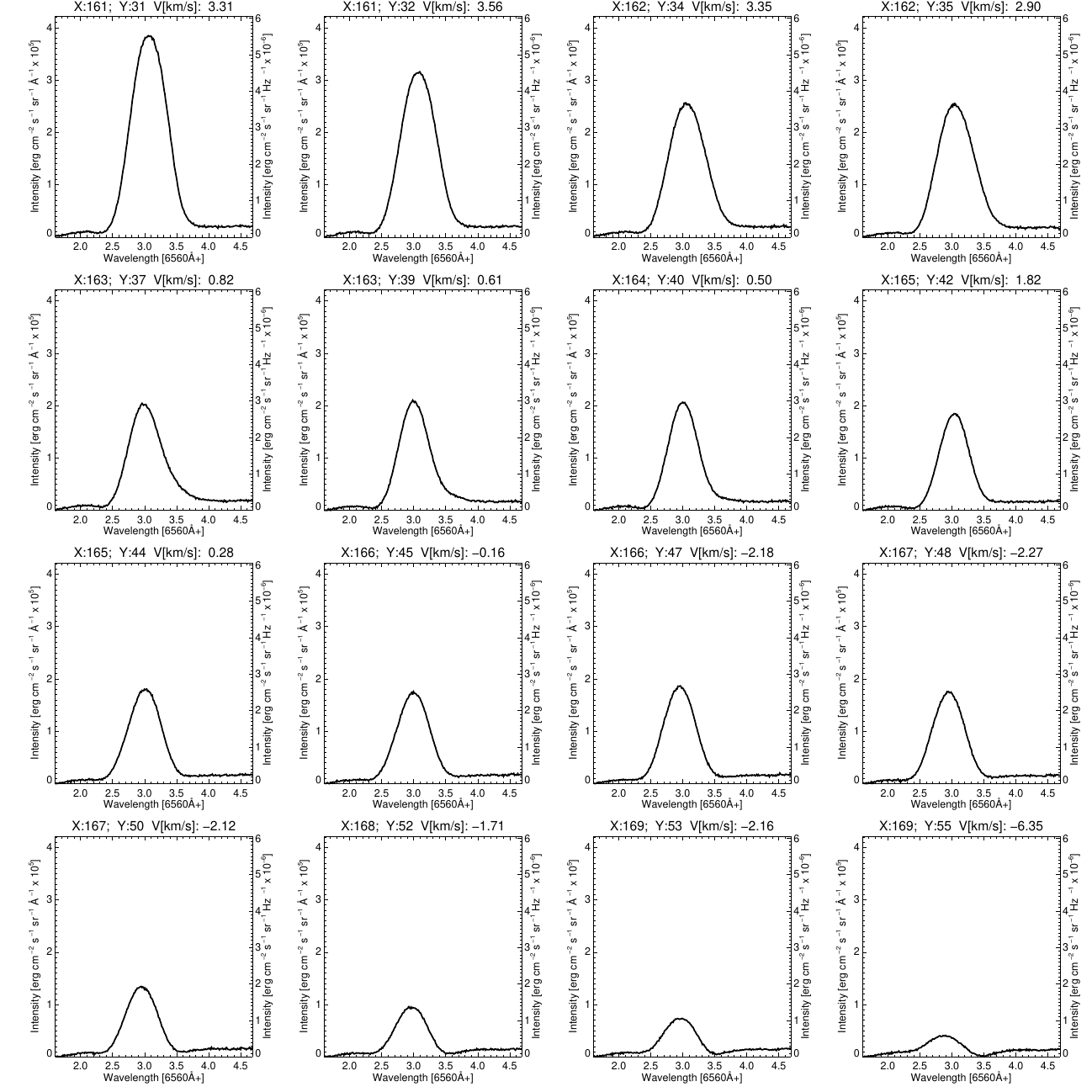}
\caption{H$\alpha$ profiles in the prominence observed with THEMIS, corresponding to the B pixels in Table~\ref{tab:mg2_ha_sel}. \label{Ha_B}}
\end{figure*}

\section{The solar limb detection}\label{ap:ap4}

{The photospheric limb can be detected using several methods.
The photospheric limb is significantly sharper than the coronal one.
In our work, we defined the limb using the pre-defined intensity threshold value.}

{Another method is, for example, to define the inflexion point, which consists of defining a limb as the extremum of the first derivative of the intensity in the function of position along the line perpendicular to the limb.
We checked  the inflexion method to define the limb of \ion{Mg}{II} continuum and H-$\alpha$ continuum. 
In Fig.~\ref{fig:img_inflection}, we show the intensity profile crossing the limb and the first derivative plot in the position function along the such profile. The strongest extremum (highlighted by a red line) defines a limb position.}

{After considering the possible errors using our threshold method compared with the inflexion method,  we decided to keep our maps. The errors for Mg~II reach one or two pixels. For H$\alpha$, due to the corrugated limb (result of the slit  scanning  the FOV and of the seeing conditions), other methods than our threshold  method would not lead to more accurate results.   Our method  takes into account  2D mapping and a threshold that is more valuable for the data that we have.}
\begin{figure*}[!htb]
\centering
\includegraphics[width=\textwidth]{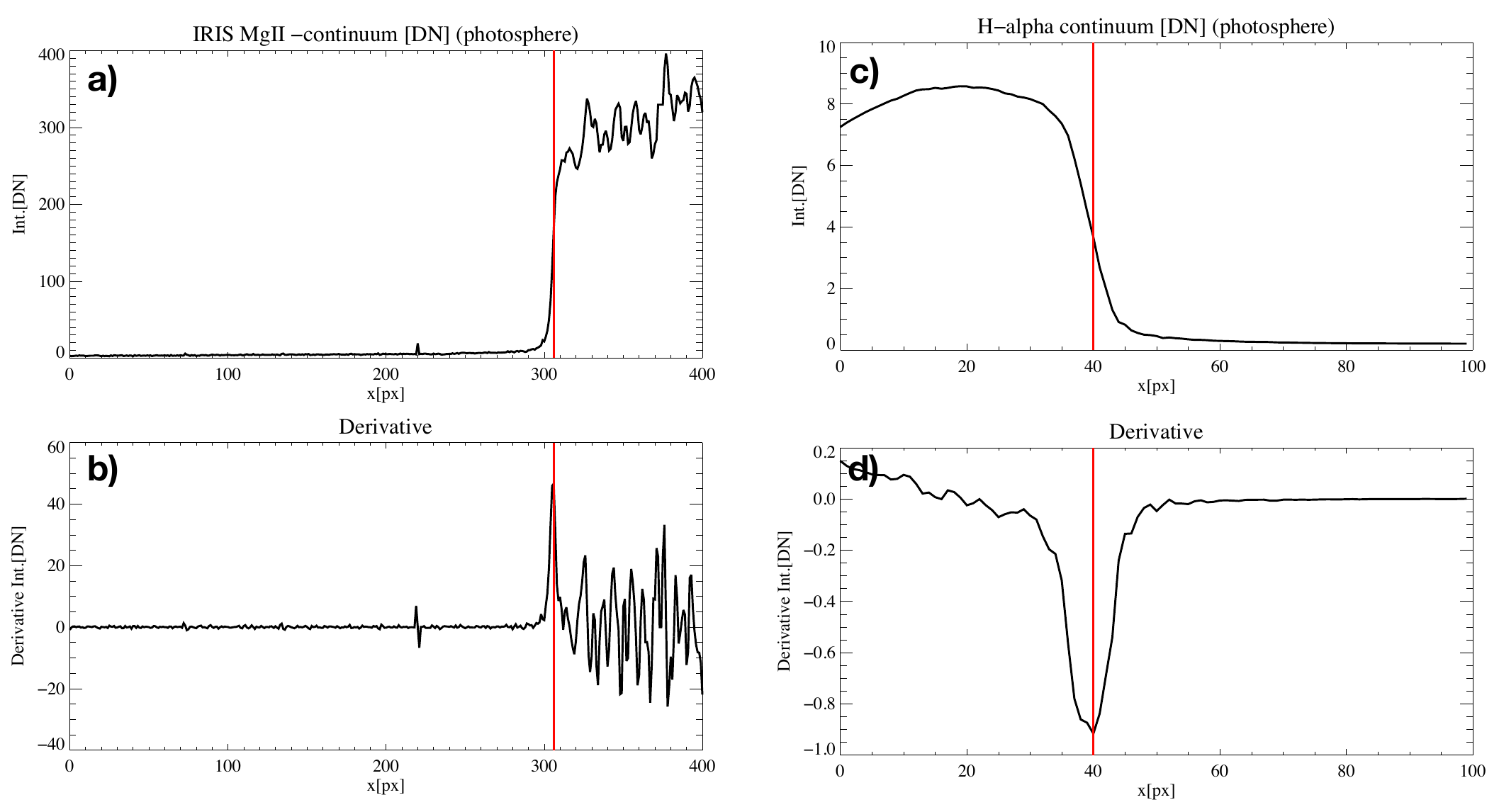}
\caption{{
Inflexion method  to define limb.  The intensity profile along straight line perpendicular to the limb in \ion{Mg}{ii} continuum (panel a) and the intensity-position derivative in the function of position (panel b) along the same profile line. The limb is defined as the strongest derivative extremum and highlighted by a red line (panel a and b).  The intensity profile parallel to the limb in H-$\alpha$ continuum (panel c) and the intensity-position derivative in the function of position (panel d). The limb is defined as the strongest derivative extremum and highlighted by a red line (panel c and d).} \label{fig:img_inflection}}
\end{figure*}

\end{appendix}
\end{document}